\documentclass{article}

\usepackage{arxiv}
\usepackage{authblk}
\usepackage{hyperref}       % hyperlinks
\usepackage{url}            % simple URL typesetting
\usepackage{booktabs}       % professional-quality tables
\usepackage{amsfonts}       % blackboard math symbols
\usepackage{nicefrac}       % compact symbols for 1/2, etc.
\usepackage{microtype}      % microtypography
\usepackage{graphicx}
\usepackage[T1]{fontenc}
\usepackage[utf8]{inputenc}
\usepackage{amsmath}
\usepackage{rotating}
\usepackage{multirow}
\usepackage{cite}
\graphicspath{ {./figures/} }

\title{Hybrid Quantum Neural Networks with Variational Quantum Regressor for Enhancing QSPR Modeling of CO$_{2}$-Capturing Amine}

\author[1]{\textbf{Hyein Cho}$^{\dagger}$}
\author[2]{\textbf{Jeonghoon Kim}$^{\dagger}$}
\author[2]{\textbf{Hocheol Lim}$^{\ast}$}
\affil[1]{The Interdisciplinary Graduate Program in Integrative Biotechnology \& Translational Medicine, Yonsei University, Incheon, Republic of Korea}
\affil[2]{Bioinformatics and Molecular Design Research Center (BMDRC), Incheon, Republic of Korea}

\begin{document}
\maketitle

\begin{flushleft}
    $^{\dagger}$ These authors contributed equally to this work. \\
    $^{\ast}$ Corresponding author: Hocheol Lim (ihc0213@yonsei.ac.kr)
\end{flushleft}

\vspace{0.5cm}

\begin{abstract}
Accurate amine property prediction is essential for optimizing CO\textsubscript{2} capture efficiency in post-combustion processes. Quantum machine learning (QML) can enhance predictive modeling by leveraging superposition, entanglement, and interference to capture complex correlations. In this study, we developed hybrid quantum neural networks (HQNN) to improve quantitative structure-property relationship modeling for CO\textsubscript{2}-capturing amines. By integrating variational quantum regressors with classical multi-layer perceptrons and graph neural networks, quantum advantages were explored in physicochemical property prediction under noiseless conditions and robustness was evaluated against quantum hardware noise using IBM quantum systems. Our results showed that HQNNs improve predictive accuracy for key solvent properties, including basicity, viscosity, boiling point, melting point, and vapor pressure. The fine-tuned and frozen pre-trained HQNN models with 9 qubits consistently achieved the highest rankings, highlighting the benefits of integrating quantum layers with pretrained classical models. Furthermore, simulations under hardware noise confirmed the robustness of HQNNs, maintaining predictive performance. Overall, these findings emphasize the potential of hybrid quantum-classical architectures in molecular modeling. As quantum hardware and QML algorithms continue to advance, practical quantum benefits in QSPR modeling and materials discovery are expected to become increasingly attainable, driven by improvements in quantum circuit design, noise mitigation, and scalable architectures.
\end{abstract}


\section{Introduction}
Carbon dioxide (CO\textsubscript{2}) is a major contributor to global warming, primarily emitted from power generation, heating, and industrial processes such as cement production, refineries, and steel manufacturing \cite{ref01_ang2016_energypolicy}. To mitigate climate change, carbon capture and storage (CCS) technologies have gained significant attention, with regulatory policies pushing industries toward emission reduction and sustainable solutions. Among these, amine-based post-combustion CO\textsubscript{2} capture has proven to be an effective and adaptable method for existing power plants  \cite{ref02_bui2018_energyenvsci}. This process uses amine solvents to absorb CO\textsubscript{2} from flue gas, making it cost-effective despite initial concerns about feasibility in the 1990s \cite{ref03_rochelle2009_science,ref04_booras1991_energy,ref05_porcheron2011_energyprocedia}. However, challenges such as high solvent volatility, equipment corrosion, and oxidative degradation increase desorption costs and pose environmental risks  \cite{ref06_chowdhury2009_energyprocedia,ref07_chowdhury2011_energyprocedia}. Addressing these issues requires identifying more efficient amine solvents, but experimentally evaluating solvent properties is costly and time-consuming. Thus, computational approaches, particularly predictive models, are essential to streamline the solvent selection process and enhance CO\textsubscript{2} capture performance.

Quantitative structure-property relationship (QSPR) is a well-established computational method in chemistry and materials science for predicting a substance's physical, chemical, or biological properties based on its molecular structure. By correlating molecular descriptors with physicochemical properties, QSPR enables efficient property estimation for new molecules, significantly reducing experimental time and cost. QSPR has been widely applied to amine solvents for CO\textsubscript{2} capture, focusing on key properties such as basicity (pKa), CO\textsubscript{2} solubility, CO\textsubscript{2} loading capacity, and absorption rate. Eshaghi Gorji \textit{et al.} predicted pKa values at various temperatures using descriptors like electronegativity and atomic charge distribution  \cite{ref08_gorji2022_chemengresdesign}, while Porcheron \textit{et al.} employed a graph-based QSPR approach to model amine thermodynamics  \cite{ref09_porcheron2013_ogst}. Khaheshi \textit{et al.} predicted the CO\textsubscript{2} absorption capacity of amine solvents using GA-MLR and LS-SVM approaches, with molecular volume, chain length, and steric hindrance as key descriptors  \cite{ref10_khaheshi2019_indengchemres}. Kuenemann \textit{et al.} demonstrated the effectiveness of cheminformatic-driven screening by applying random forests and neural networks to predict CO\textsubscript{2} absorption capacity  \cite{ref11_kuenemann2017_molinf}. Despite these efforts, building more accurate models with limited experimental databases remains a critical challenge. Addressing this limitation is essential for advancing CCS technology and improving solvent performance.

Quantum computing, based on superposition and entanglement, has transformative potential in chemistry, optimization, and materials science \cite{ref12_biamonte2017_nature,ref13_mcardle2020_revmodphys,ref14_nannicini2019_physicalreviewe,ref15_hirai2024_scientificreports,ref16_ray2024_qce,ref17_hirai2023_arxiv}. Quantum machine learning (QML), the quantum counterpart to classical machine learning, offers advantages in handling quantum data and solving complex problems beyond classical capabilities. As a subset of QML, quantum neural networks (QNNs) resemble architectures of neural networks with the capability of carrying out universal quantum computation  \cite{ref18_beer2020_naturecommunications}. Three main types of QNN have been proposed, each addressing different aspects of quantum computation and learning. First, the dissipative model uses open quantum systems, leveraging dissipation to guide the system toward stable states that encode solutions to optimization problems. This model extends traditional feedforward networks by employing unitary operators and discarding qubits in layers once their information has dissipated into subsequent layers to manage qubit counts as depth increases  \cite{ref18_beer2020_naturecommunications}. Second, the iterative QNN model uses parameterized quantum circuits with unitary transformations evolving iteratively. These circuits are optimized via gradient-based or variational methods, enabling efficient learning and robust classification of quantum data  \cite{ref19_farhi2018_arxiv}. Third, quantum convolutional neural networks (QCNNs) apply quantum gates in convolutional layers followed by pooling operations. This approach reduces state complexity by measuring qubits at each stage, lowering data dimensionality while preserving essential features  \cite{ref20_cong2019_naturephysics}. These QML algorithms enhance clustering, classification, and computational efficiency in data analysis  \cite{ref21_tomar2025_arxiv} and support quantum transfer learning to improve generalization and reduce overfitting  \cite{ref15_hirai2024_scientificreports}. However, current noisy intermediate-scale quantum (NISQ) devices face limitations, such as restricted qubit counts, limited connectivity, and hardware noise, which constrain quantum advantage. Despite these challenges, QML has already shown improved performance and speedups in practical applications, even with limited data \cite{ref22_arute2019_nature,ref23_zhong2020_science,ref24_montanaro2016_npjquantuminf}.

 In this study, we developed hybrid quantum neural networks (HQNN) with a variational quantum regressor (VQR) to enhance QSPR modeling for CO\textsubscript{2}-capturing amine solvents. First, we collected five key properties of amines, such as basicity, viscosity, boiling point, melting point, and vapor pressure, which are critical for CO\textsubscript{2} capture and for addressing limitations in existing solvents. Next, we built classical QSPR models using multi-layer perceptrons (MLP) and graph neural networks (GNN) and extended them into HQNN to leverage quantum advantages for improved performance. We then evaluated various training strategies to identify the most effective approach for maximizing quantum advantages under noiseless conditions. Finally, we assessed the impact of hardware noise in NISQ devices through noisy simulations. Our results demonstrate that quantum-enhanced modeling improves QSPR performance, enabling more accurate predictions of amine solvent properties and supporting the discovery of efficient amines for CO\textsubscript{2} capture.

\section{Methods}
Experimental data for basicity (pKa) were collected from the literature \cite{ref25_yu2010_jcheminf,ref26_baltruschat2020_f1000research,ref27_pan2021_jcheminf,ref28_mansouri2019_jcheminf}, selecting only the nitrogen pKa values and using the average for amines with multiple nitrogens. Viscosity data were sourced from Chew \textit{et al.}  \cite{ref29_chew2024_jcheminf} and log-scale transformed. Boiling point, melting point, and vapor pressure data were retrieved from EPI Suite 4.11  \cite{ref30_epa2012_epa}, with additional boiling point data manually added from other sources  \cite{ref31_kazakov2012_ijthermophys,ref32_williams2017_jcheminf}. The number of data points and their minimum-maximum ranges are summarized in Table~\ref{tab:dataset_summary}.

\begin{table}[ht!]
    \centering
    \caption{Summary of datasets used in this work.}
    \renewcommand{\arraystretch}{1.3} % Adjust row spacing
    \begin{tabular}{@{}c c c c c@{}}
        \toprule
        \textbf{Task} & \textbf{Unit} & \textbf{All set} & \textbf{Minimum} & \textbf{Maximum} \\
        \specialrule{1.5pt}{0pt}{0pt}
        Basicity (pKa) & - & 6469 & -2.86 & 11.94 \\
        Viscosity & $\log_{10}(\eta)$ (cP) & 3582 & -1.00 & 1.42 \\
        Vapor Pressure & $\log_{10}(P)$ (mmHg) & 2945 & -20.74 & 7.00 \\
        Boiling Point & $T_{\mathrm{B}}$ (°C) & 23,044 & -268.93 & 5900 \\
        Melting Point & $T_{\mathrm{M}}$ (°C) & 9721 & -219.61 & 3410 \\
        \bottomrule
    \end{tabular}
    \label{tab:dataset_summary}
\end{table}

\subsection{Classical Neural Networks (Classical NN)}

All datasets were split into training and validation sets using a fixed random seed for 5-fold cross-validation. A stratified split was applied to balance the scaffolds and the \textit{y}-values across the quintiles. The scaffold split was performed with MurckoScaffold from RDKit to assign a scaffold index to each molecule  \cite{ref33_landrum2013_rdkit}, while the \textit{y}-values were distributed into the quintiles based on their magnitude. Grid search was used to optimize hyperparameters, as shown in Table S1, and performance was evaluated using R\textsuperscript{2} and mean absolute error (MAE).

\subsubsection{Molecular Fingerprints and Multi-Layer Perceptron (MLP)}

Molecular fingerprints numerically represent molecular features, with 11 types used: MACCS, Avalon, ECFP6, FCFP4, PCFP, Extended, Morgan, rDesc, rPair, rTorsion, and Standard. Molecular access system (MACCS) fingerprints use 166-bit keys for predefined substructures  \cite{ref34_durant2002_jcheminf}. Avalon fingerprints employ a 1024-bit path-based vector for molecular searches  \cite{ref35_gedeck2006_jcheminf}. Extended-connectivity fingerprints with a radius of 6 (ECFP6) and Functional-class fingerprints with a radius of 4 (FCFP4) use circular atom neighborhoods with 1024-bit keys, capturing substructure and pharmacophoric features, respectively  \cite{ref33_landrum2013_rdkit,ref36_rogers2010_jcheminf}. The PubChem fingerprints (PCFP) encode substructures for chemical similarity  \cite{ref37_bolton2008_annrepcompchem}, while the Extended fingerprint considers rings and atomic properties within a 1024-bit vector. The Morgan fingerprint uses atom neighborhoods and the Morgan algorithm to represent chemical structures  \cite{ref38_morgan1965_jchemdoc,ref39_tailor2021_arxiv}. rDesc, rPair, and rTorsion capture ring systems, ring pairs, and torsional relationships, providing structural insights. The Standard fingerprint uses a 1024-bit vector to represent structural fragments for similarity searches  \cite{ref40_carhart1985_jcheminf}. To enhance feature diversity and improve deep learning performance, we concatenated MACCS with six other fingerprints (Avalon, ECFP6, Extended, FCFP4, PCFP, and Standard).

These fingerprints were trained using a multi-layer perceptron (MLP), a widely used feedforward neural network with multiple layers \cite{ref41_goodfellow2016_deeplearning}. Input signals pass through hidden layers in one direction without loops, with neurons fully connected to adjacent layers via weighted connections. The training process adjusts weights and biases through backpropagation, minimizing a loss function by computing gradients and updating parameters with optimization algorithms. We performed a grid search on six hidden layer configurations, as shown in Table S1, using ReLU  \cite{ref42_agarap2018_arxiv} as the activation function and the Adam optimizer  \cite{ref43_kingma2014_arxiv}. The \textit{y}-values were min-max scaled for MLP training, ensuring compatibility with the hybrid quantum neural network. During evaluation, the scaled predictions were converted back to their original real values for performance assessment.

\subsubsection{Molecular Graph and Graph Neural Networks (GNN)}

Molecular structures, with their 3-dimensional nature, can be represented as molecular graphs consisting of nodes and edges. Nodes represent atoms with features like element (44-dimensional), degree (6-dimensional), valence (6-dimensional), hydrogen count (5-dimensional), and aromaticity (2-dimensional). Edges represent bonds with features such as bond type (4-dimensional), conjugation (2-dimensional), stereoisomerism (3-dimensional), and ring presence (2-dimensional). These graphs capture both structural and electronic properties. Graph neural networks (GNNs) were employed using DMPNN, EGC, GAT, GCN, and TCN. Directed message-passing neural network (DMPNN) enhances message passing with bond-specific information  \cite{ref44_dai2016_icml}. An efficient graph convolutional network (EGC) efficiently utilizes spatial data with isotropic message passing. Graph attention transformer (GAT) applies attention mechanisms to weigh neighborhood interactions  \cite{ref45_brody2021_arxiv}. Graph convolutional network (GCN) aggregates data to capture local and global patterns while preventing over-smoothing with residual connections  \cite{ref46_chen2020_icml}. A transformer convolutional network (TCN) uses transformer-based attention for complex structural representations in large datasets  \cite{ref47_shi2020_arxiv}. A grid search was conducted on two hyperparameters for DMPNN and GCN, and three for EGC, GAT, and TCN, as detailed in Table S1. We also used ReLU  \cite{ref42_agarap2018_arxiv} as the activation function and the Adam optimizer  \cite{ref43_kingma2014_arxiv}. Unlike MLP models, GNNs were trained with unnormalized real values to preserve accurate node relationships.

\subsection{Hybrid Quantum Neural Networks (HQNN)}

Quantum computing leverages quantum physics for efficient computation. A single qubit is a unit vector in 2-dimensional Hilbert space in Equation~\ref{eqn1},

\begin{equation}
\label{eqn1}
\left\vert \psi \right\rangle = \alpha \left\vert 0 \right\rangle + \beta \left\vert 1 \right\rangle
\end{equation}

where  \( \vert 0\rangle\) and \( \vert 1\rangle\) represent classical bits 0 and 1. An \textit{n}-qubit state exists in a 2\textsuperscript{\textit{n}}-dimensional space (Equation~\ref{eqn2}).

\begin{equation}
\label{eqn2}
\left\vert \psi \right\rangle = \sum_{x \in \{0,1\}^n} \alpha_{x} \left\vert x \right\rangle
\end{equation}

A measurement of an \textit{n}-qubit state yields one of the classical bit strings $x$ with probability $ \vert \alpha_{x} \vert^{2} $. Quantum circuits start from an initial state and apply operations, such as \textit{H}, \textit{X}, \textit{Y}, \textit{Z}, and CNOT, to generate a final state. Parameterized gates are used to shape the output distribution  \cite{ref48_nielsen2010_quantumcomp}.

 Variational quantum regressor (VQR) optimizes parameters to minimize the mean squared error (MSE) between true and predicted values. The architecture comprises an encoder, an ansatz, and a decoder. The encoding process applies to the \textit{R\textsubscript{y}} rotation gate with the angle for the scaled attribute $x$. The arctangent function ensures that values outside the (-1, 1) can be converted to a unique angle after scaling. Each attribute is encoded into a single qubit from the \textit{n}-qubit quantum models. The ansatz is implemented with a circuit of \textit{R\textsubscript{y}} rotation gates and CNOT gates in a linear configuration, where the learnable parameters correspond to the rotation angles of the \textit{R\textsubscript{y}} gates. The models were constructed by stacking $d$ blocks of one-depth two-qubit layers, where $d$ denotes the circuit depth. The decoder generates predictions by calculating the expectation value of the observable quantum state produced by the encoder-ansatz circuit. The expectation value is computed using the $\sigma_{z}$ operator, summed across all qubits. 

The architecture integrates classical NNs (MLP and GNN) with a quantum layer for final state transformation. Based on classical performance, either MLP or GNN is attached to the input layer, producing embeddings for the quantum encoder. The classical embeddings are utilized to make predictions by applying the transformation defined in Equations~\ref{eqn3}, \ref{eqn4}, \ref{eqn5}, \ref{eqn6}, and \ref{eqn7}.

\begin{equation}
\label{eqn3}
\mathrm{Embed}_{\text{classical NN}} = f_{\text{classical NN}}(x; \theta_{\text{classical NN}})
\end{equation}

\begin{equation}
\label{eqn4}
\theta_{i} = \arctan \left( \mathrm{Embed}_{i} \right) + \frac{\pi}{2}
\end{equation}

The classical embeddings are computed as described in Equation~\ref{eqn3}. These embeddings are then used by the encoder to determine the \textit{R\textsubscript{y}} gate rotation angles (Equation~\ref{eqn4}).

\begin{equation}
\label{eqn5}
V(\theta) = \prod_{j} \left[ R_{y}(\theta_{j}) \cdot \mathrm{CNOT}_{j, j+1} \right]
\end{equation}

\begin{equation}
\label{eqn6}
\left\vert \psi_{\text{ansatz}} \right\rangle = V^{d}(\theta) \left\vert \psi_{\text{enc}} \right\rangle
\end{equation}

\begin{equation}
\label{eqn7}
y_{\text{pred}} = \sum_{i} \left\langle \psi_{\text{ansatz}} \vert \sigma_{z}^{i} \vert \psi_{\text{ansatz}} \right\rangle
\end{equation}

The resulting angles are passed to the variational quantum circuit (VQC), which applies a feature map followed by an ansatz (Equation~\ref{eqn5}). The complete ansatz circuit after layers is defined in Equation~\ref{eqn6}. The decoder subsequently generates predictions by calculating the expectation value of the Pauli-Z projection across all qubits (Equation~\ref{eqn7}). To evaluate training performance, HQNN compares three strategies: training from scratch, finetuning, and freezing pre-trained weights. The implementation is built using Python 3.10 and the PennyLane framework 0.38.0 with PyTorch 2.4.0 integration  \cite{ref49_bergholm2018_arxiv}.

\begin{figure}[ht!]
  \centering
  \includegraphics[width=1.0\textwidth]{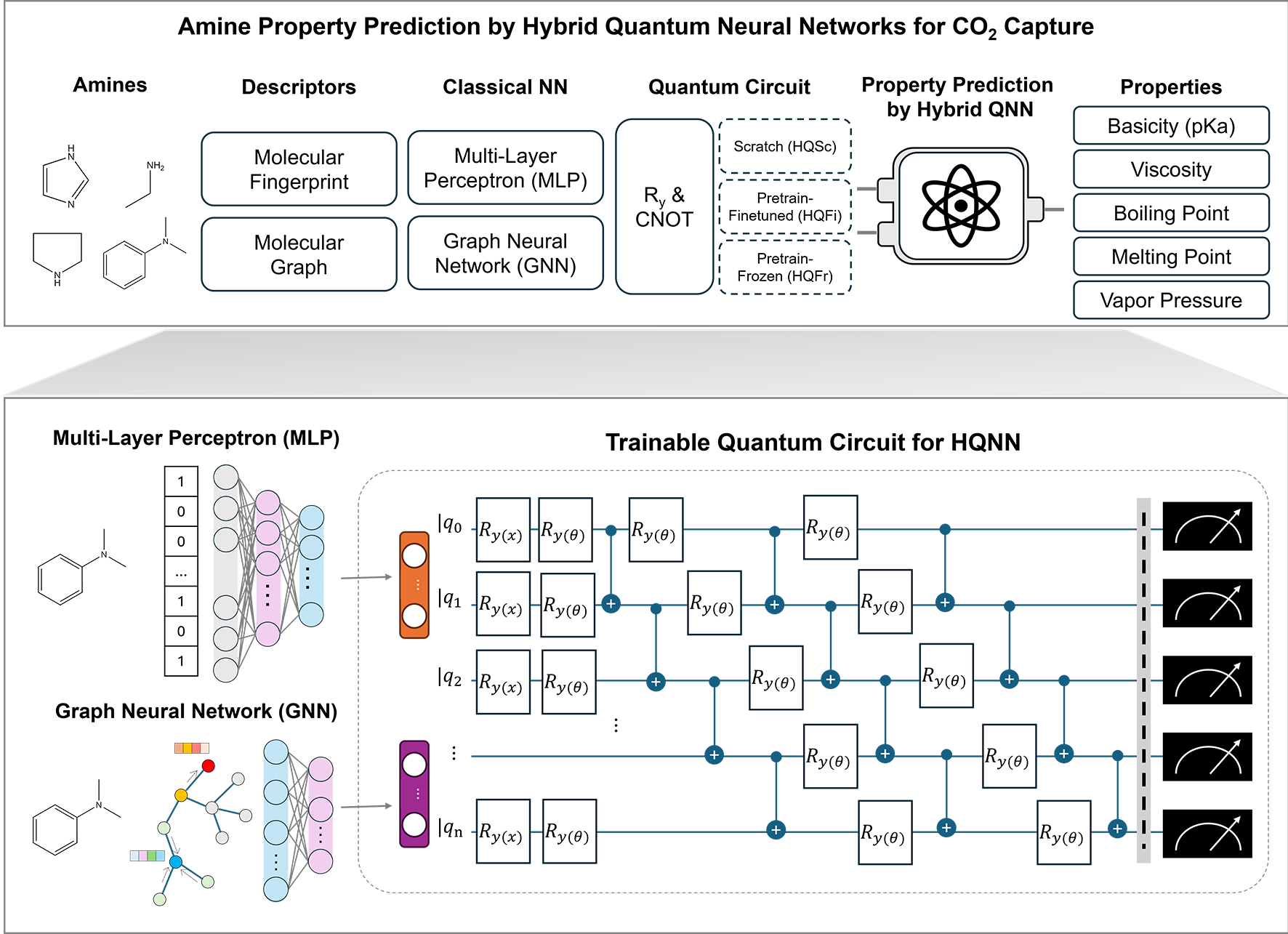}
  \caption{Overview of HQNN with VQR for QSPR Modeling of CO$_{2}$-Capturing Amine Properties}
  \label{fig:fig1_graphic_workflow}
\end{figure}

\subsection{Hardware Noisy Simulation}

To evaluate the reliability and robustness of HQNN in a hardware environment, we considered four noise sources: depolarizing channels, amplitude damping, phase damping, and readout errors. The two-qubit depolarizing channels were determined using the lowest two-qubit error rate. In contrast, the one-qubit depolarizing channels were calculated with the median error rate of the SX gate. The two-qubit and SX gate error rates represent errors occurring during the execution of two-qubit and SX gates, with lower values indicating higher reliability. Amplitude damping and phase damping were calculated using relaxation time (T\textsubscript{1}), decoherence time (T\textsubscript{2}), and gate time. A longer relaxation time improves resistance to noise, while a longer decoherence time allows for more complex quantum computations. Gate time refers to the duration required to perform a specific gate operation; shorter gate times enhance computation speed and reduce errors from decoherence. Additionally, readout error occurs during qubit state measurement, with lower values indicating higher accuracy. The noise parameters for the noisy simulation were obtained from seven real quantum hardware devices (IBM-Fez, IBM-Marrakesh, IBM-Torino, IBM-Yonsei, IBM-Brisbane, IBM-Brussels, and IBM-Strasbourg) to evaluate performance consistency (Table S8).

\section{Results}

Amine-based CO\textsubscript{2} capture relies on accurate solvent property prediction to improve efficiency and reduce costs. To enhance predictive performance, we developed a hybrid quantum neural network (HQNN) by integrating the best-performing classical neural networks (NN) with a quantum neural network (QNN), as shown in Figure~\ref{fig:fig1_graphic_workflow}. Our model predicts five key properties of CO\textsubscript{2}-capturing amines, including basicity (pKa), viscosity, vapor pressure, boiling point, and melting point, summarized in Table~\ref{tab:dataset_summary}. These properties are critical for selecting effective CO\textsubscript{2} absorbents, which require a high reaction rate, substantial CO\textsubscript{2} absorption capacity, efficient renewable energy utilization, and environmental compatibility.

\begin{table}[ht!]
    \centering
    \caption{The Relative Performance (\%) of HQNN Models for Physicochemical Properties of CO\textsubscript{2}-Capturing Amines}
    \label{tab:hqnn_performance}
    \renewcommand{\arraystretch}{1.3}  % Adjust row spacing
    \setlength{\tabcolsep}{4pt}  % Adjust column spacing
    
    \begin{tabular}{@{}cccccccccccccc@{}}
        \toprule
        \multirow{2}{*}{\textbf{Property}} & \multirow{2}{*}{\textbf{Model}} 
          & \multicolumn{2}{r}{\textbf{HQSc (4Q)}} 
          & \multicolumn{2}{r}{\textbf{HQSc (9Q)}} 
          & \multicolumn{2}{r}{\textbf{HQFi (4Q)}} 
          & \multicolumn{2}{r}{\textbf{HQFi (9Q)}} 
          & \multicolumn{2}{r}{\textbf{HQFr (4Q)}} 
          & \multicolumn{2}{r}{\textbf{HQFr (9Q)}} \\
          \cmidrule(lr){3-14}
        % \cmidrule(lr){3-14} \cmidrule(lr){5-6} \cmidrule(lr){7-8} \cmidrule(lr){9-10} \cmidrule(lr){11-12} \cmidrule(lr){13-14}
         & & \textbf{R$^2$} & \textbf{MAE} & \textbf{R$^2$} & \textbf{MAE} 
           & \textbf{R$^2$} & \textbf{MAE} & \textbf{R$^2$} & \textbf{MAE} 
           & \textbf{R$^2$} & \textbf{MAE} & \textbf{R$^2$} & \textbf{MAE} \\
        \specialrule{1.5pt}{0pt}{0pt}
        
        % Basicity (pKa)
        Basicity (pKa) & HQMLP 
           & 0.07  & 6.19  & -0.03 & 4.35  & 0.33  & 7.55  
           & 0.13  & 6.30  & 0.12  & 2.22  & 0.12  & 2.16  \\
        \midrule
        
        % Viscosity
        \multirow{2}{*}{Viscosity} 
         & HQMLP 
           & -0.48 & -0.11 & -1.36 & -0.68 & -0.78 & 0.34  
           & -0.49 & 0.34  & -0.64 & 0.23  & -0.56 & 0.23  \\
         & HQGNN  
           & -0.75 & -0.34 & 0.41  & 0.73  & -0.89 & -0.23 
           & -0.23 & 0.17  & -0.71 & -0.68 & 0.04  & 0.68  \\
        \midrule
        
        % Vapor Pressure
        Vapor Pressure & HQMLP 
           & -3.33 & -10.45 & -1.16 & -5.39 & -0.25 & -0.41  
           & -0.44 & 1.69   & -0.41 & -5.13  & -0.35 & -3.73  \\
        \midrule
        
        % Boiling Point
        Boiling Point & HQMLP 
           & -0.19 & -1.73  & 0.39  & -8.61  & -0.16 & -3.93  
           & -0.49 & -8.66  & 0.24  & -3.26  & 0.21  & -1.66  \\
        \midrule
        
        % Melting Point
        Melting Point & HQGNN 
           & -0.85 & 1.25  & -2.63 & -0.11  & 1.26  & -1.05  
           & 3.52  & 0.35  & -1.17 & -1.33  & 2.43  & 0.02   \\
        \bottomrule
    \end{tabular}
\end{table}

To identify the most effective classical NN, we constructed QSPR models using molecular fingerprints with multi-layer perceptrons (MLP) and molecular graphs with graph neural networks (GNN), capturing both structural and electronic characteristics of solvents. These models were optimized through a hyperparameter tuning process using grid search and 5-fold cross-validation, with the procedure summarized in Table S1 and the results in Table S2. We then evaluated the predictive performance of classical NNs for each property using R\textsuperscript{2} and MAE across MLP, GNN, and their combination model, with performance metrics detailed in Tables S3-S7.

\subsection{Hybrid Quantum Neural Networks (HQNN)}

Building on the best-performing classical NN, we developed HQNN to enhance predictive performance by integrating quantum computation. HQNN utilizes either an MLP or GNN for initial feature extraction, followed by a quantum layer based on Hirai’s architecture  \cite{ref15_hirai2024_scientificreports}, which employs \textit{R\textsubscript{y}} rotation and CNOT gates. The quantum layer serves as the final transformation stage, leveraging quantum computation to enhance predictive accuracy while capturing complex molecular interactions.

 To systematically evaluate HQNN, we implemented three training strategies, as illustrated in Figure~\ref{fig:fig1_graphic_workflow}. The first approach, HQNN from Scratch (HQSc), follows the architecture of MLP or GNN but initializes all weights randomly instead of using pre-trained values. This allows the model to learn task-specific features without bias from prior training. The second approach, HQNN Pretrain-Finetuned (HQFi), attaches a quantum layer to a pre-trained classical model and updates all weights, including those in the classical MLP or GNN layers, during training. This strategy enables the network to adapt to the target task while retaining knowledge encoded in the pre-trained weights. The third approach, HQNN Pretrain-Frozen (HQFr), treats the pre-trained classical model as a fixed feature extractor, training only the quantum layer. This preserves the learned representations of the classical model while optimizing the quantum transformation layer.

The HQNN architectures were implemented with either 4 or 9 qubits, following the requirements set by Zaman \textit{et al.} for maintaining a square number of qubits in quantum convolutional structures such as Quanvolutional Neural Networks (QuanNN) and Quantum ResNet (QResNet) \cite{ref50_zaman2024_arxiv}. Increasing the qubit count allows for shallower circuit depths, improving feasibility on noisy intermediate-scale quantum (NISQ) devices while ensuring compatibility with quantum hardware constraints. By designing architectures with different qubit configurations, we aimed to balance computational efficiency with the limitations of current quantum devices.

\begin{figure}[ht!]
  \centering
  \includegraphics[width=1.0\textwidth]{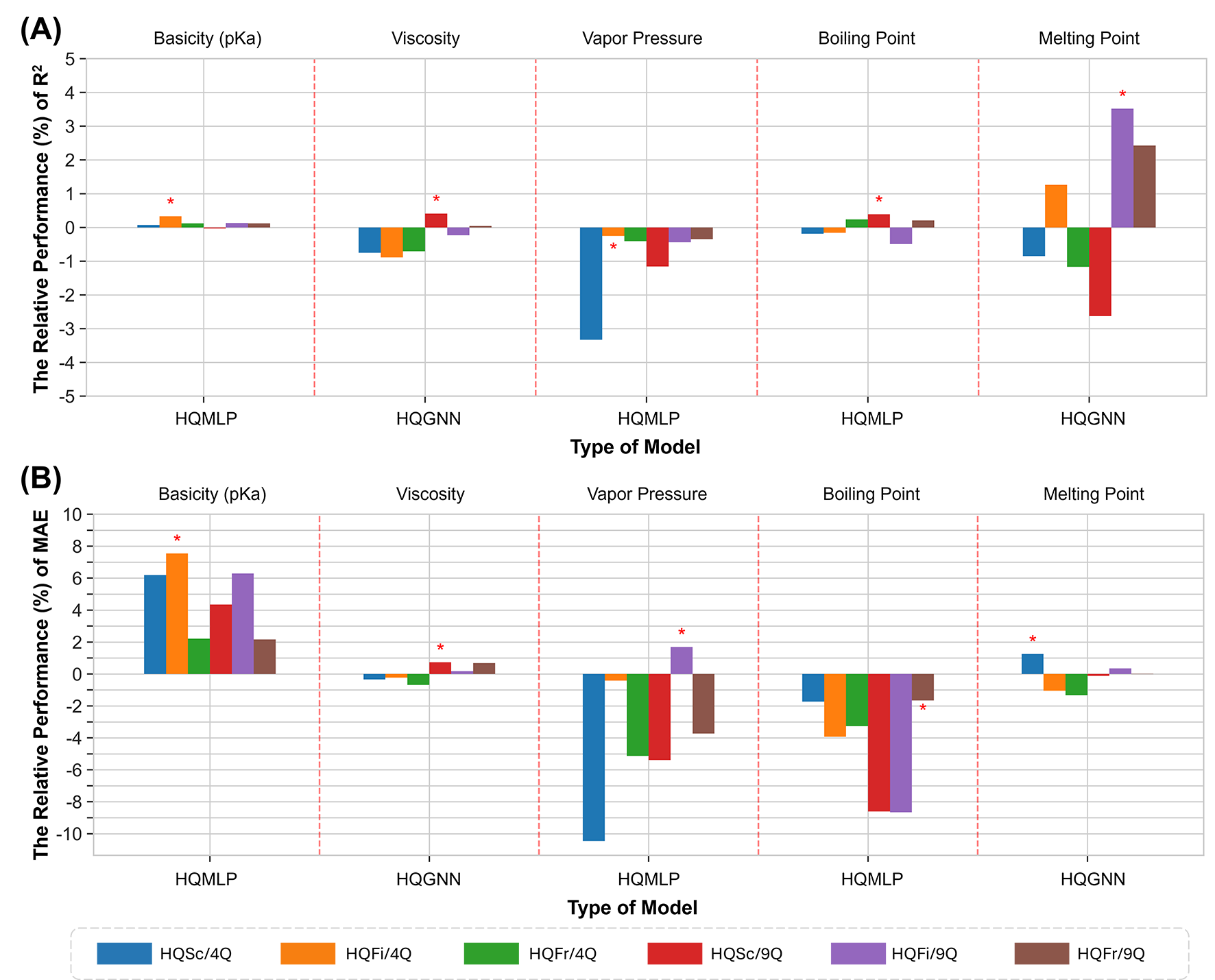}
    \caption{The relative performance of classical NNs and HQNNs for five physicochemical properties. (A) Bar plot of relative $\text{R}^2$ performance for predicting basicity, viscosity, vapor pressure, boiling point, and melting point. (B) Bar plot of relative MAE performance. Models were evaluated using 5-fold cross-validation across three training strategies (HQSc, HQFi, HQFr) with 4Q and 9Q.}
    
  \label{fig:fig2_hqnn_performance}
\end{figure}

By systematically comparing these training strategies, we assessed the impact of pre-trained knowledge versus full-network training in hybrid quantum-classical models. This evaluation provides insights into the optimal integration of classical and quantum machine learning paradigms for QSPR applications. The overall quantum advantage is illustrated in Figure~\ref{fig:fig2_hqnn_performance}, while relative performance improvements across evaluation metrics are summarized in Table~\ref{tab:hqnn_performance}.

\subsection{Prediction of Key Physicochemical Properties}

The basicity of amines strongly correlates with CO\textsubscript{2} solubility and can serve as a proxy for cyclic capacity. Higher basicity values lead to more stable carbamate formation, which reduces CO\textsubscript{2} regeneration efficiency. This inverse relationship highlights that as carbamate stability increases, CO\textsubscript{2} desorption efficiency decreases, resulting in higher CO\textsubscript{2} loading capacities within the amine solution  \cite{ref51_rezaei2016_jngse,ref52_bernhardsen2017_ijggc}.Among classical NNs, MLP outperformed GNN, achieving an  R\textsuperscript{2} of 0.9082 and an MAE of 0.3746 (Table S3). Although combining MLP and GNN was expected to enhance representation by integrating numerical and graph features, it resulted in lower performance than either model alone. For HQNNs, HQMLP with 4 qubits and HQFi training achieved the best performance, with an  R\textsuperscript{2}  of 0.9112 and an MAE of 0.3463 (Table S3). HQNN architectures consistently demonstrated quantum advantages across all training strategies, suggesting that this task benefits from quantum-enhanced learning. The HQFi/4Q model showed the highest quantum gain, improving R\textsuperscript{2} by 0.33$\%$ and reducing MAE by 7.55$\%$ (Table~\ref{tab:hqnn_performance}).

Viscosity significantly influences CO\textsubscript{2} solubility by affecting mass transfer rates. Increased viscosity reduces absorption rates and capacity, necessitating an optimal balance between viscosity and absorption efficiency  \cite{ref53_yuan2017_energyprocedia,ref54_li2014_ijggc}. Among classical models, the DMPNN achieved the highest R\textsuperscript{2} of 0.7304 and an MAE of 0.1786, while MLP showed a slightly lower MAE of 0.1771. For HQNNs, HQSc/9Q using DMPNN exhibited the best performance in terms of  R\textsuperscript{2}, while other HQGNN models had lower MAE than their classical counterparts. Although HQMLP improved MAE except for HQSc/4Q and HQSc/9Q, it did not significantly enhance R\textsuperscript{2 }(Table S4). As a result, HQGNN training with HQSc/9Q and HQFr/9Q achieved a quantum advantage in both R\textsuperscript{2} and MAE, indicating that higher qubit counts may enhance viscosity prediction accuracy (Figure~\ref{fig:fig2_hqnn_performance}).

\begin{figure}[ht!]
  \centering
  \includegraphics[width=1.0\textwidth]{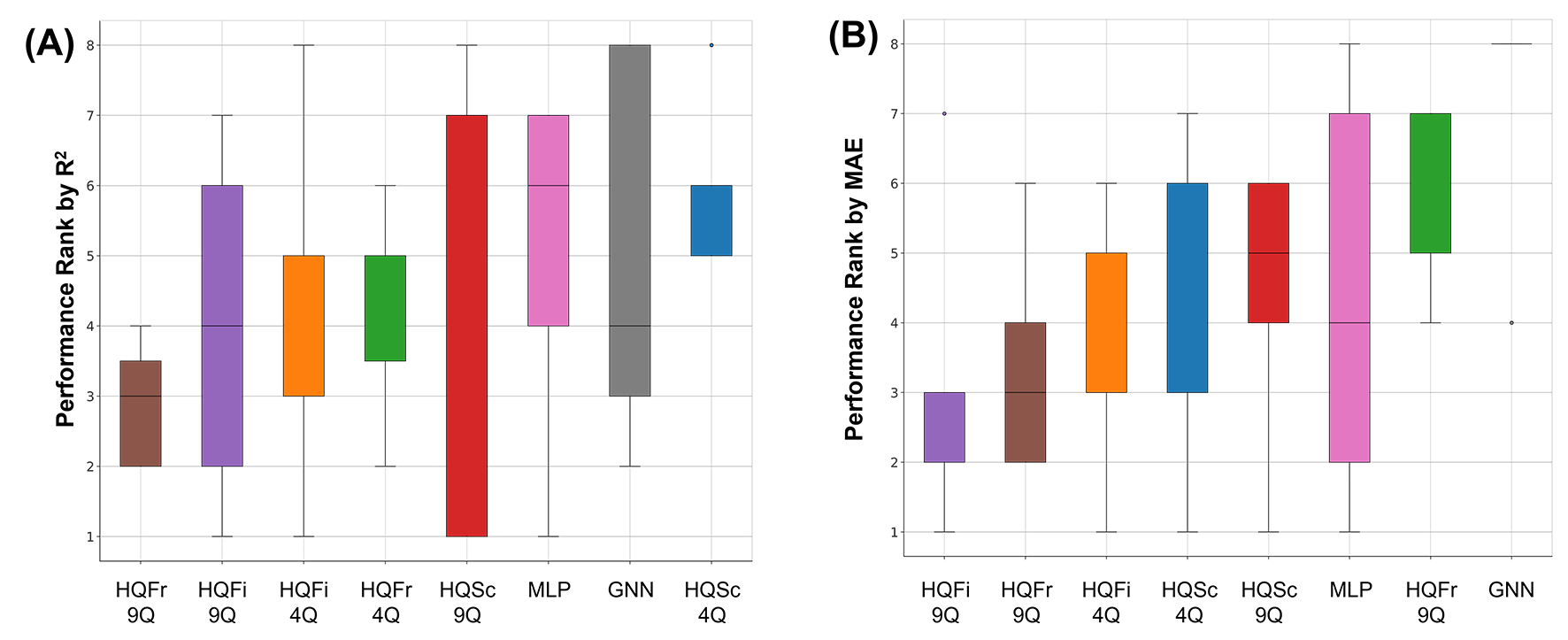}
  \caption{The box plot comparison rank of classical NNs and HQNNs. (A) Box plot comparison rank of models by $\text{R}^2$ metric in the test set. (B) Box plot comparison rank of models by MAE metric in the test set. Models were evaluated using 5-fold cross-validation across three training strategies (HQSc, HQFi, HQFr) with 4Q and 9Q. The models are ordered from left to right based on increasing average rankings.}

  \label{fig:fig3_boxplot_ranking}
\end{figure}

Vapor pressure plays a crucial role in CO\textsubscript{2} capture, affecting solvent recovery, energy consumption, and operational efficiency. It provides key insights into energy requirements for solvent regeneration and helps mitigate losses due to evaporation  \cite{ref55_wu2013_exptthermfluid}. Among classical NNs, MLP performed best, achieving an R\textsuperscript{2} of 0.8860 and an MAE of 0.6868. Despite expectations, combining MLP and DMPNN did not yield performance gains. Among HQNNs, HQFi/4Q achieved the highest R\textsuperscript{2} of 0.8838, but it remained slightly below that of the classical MLP. However, HQFi/9Q attained the lowest MAE of 0.6752, outperforming both classical models and other HQNN configurations. Unlike other properties, integrating the best classical NN with VQR vapor pressure prediction exhibited the most substantial performance decline, with HQSc/4Q showing the weakest results in both R\textsuperscript{2} and MAE (Table S5).

The boiling point of amine solvents influences CO\textsubscript{2} capture efficiency by affecting stability and volatility. Higher boiling points generally reduce solvent losses, but balancing this property with absorption efficiency is crucial for optimal performance  \cite{ref56_freeman2010_ijggc,ref57_yang2017_chemrev}. Based on grid search results, MLP demonstrated the best performance among classical NNs, leading to its selection for HQNN construction. However, HQNNs did not improve MAE across any training strategy. The HQFr/4Q, HQSc/9Q, and HQFr/9Q models achieved higher R\textsuperscript{2} than MLP, with HQSc/9Q attaining the highest R\textsuperscript{2} of 0.8788 among all architectures. The HQFr/4Q model also exhibited an improved R\textsuperscript{2} of 0.8770 (Table S6).

Melting point is a critical physicochemical property affecting CO\textsubscript{2} solubility, operational flexibility, and overall capture capacity. Lower melting points often enhance solubility, improving CO\textsubscript{2} absorption and mass transfer efficiency  \cite{ref58_katritzky2001_crystgrowthdes}. Among classical NNs, the EGC model achieved the best performance with an R\textsuperscript{2} of 0.7619 and an MAE of 35.4447. The combined EGC-MLP model showed diminished performance, indicating that increased input features did not necessarily enhance training results. HQNNs were constructed using EGC architecture, with HQFi/9Q demonstrating the best performance (R\textsuperscript{2} of 0.7887 and MAE of 35.3202). HQFi/4Q, HQFi/9Q, and HQFr/9Q outperformed classical models in R\textsuperscript{2}, while HQSc/4Q achieved lower MAE values, demonstrating an advantage over classical architectures. The best-performing HQNN, HQFi/9Q, improved R\textsuperscript{2} by 3.52$\%$ and MAE by 0.35$\%$ (Table S7).

Overall, the impact of HQNN architectures varied across the physicochemical properties. Among them, basicity exhibited the most significant improvement, with a 7.55 $\%$ reduction in MAE, whereas vapor pressure experienced the greatest decline with quantum implementation. Notably, combining MLP with GNN did not lead to any performance improvement, suggesting that increasing input complexity does not necessarily enhance property prediction. These findings indicate that the effectiveness of quantum implementation is property-dependent and underscore the need for carefully selecting both model architectures and quantum strategies to maximize predictive performance.

\begin{table}[ht!]
    \centering
    % \small
    \caption{Prediction Performance of HQFi for Basicity (pKa) Across IBM Quantum Hardware Noise}
    \label{tab:noise_pka}
    \setlength{\tabcolsep}{4pt}
    \renewcommand{\arraystretch}{1.3} % Adjust row height for better readability
    \begin{tabular}{ccccccc}
        \toprule
        \textbf{Task} & \textbf{Model} & \textbf{Noise} & \textbf{Train R$^2$} & \textbf{Train MAE} & \textbf{Test R$^2$} & \textbf{Test MAE} \\
        \specialrule{1.5pt}{0pt}{0pt}
        \multirow{9}{*}{\shortstack{Basicity\\(pKa)}} & \multirow{9}{*}{\shortstack{HQMLP \\ (HQFi/4Q)}} & Noiseless & 0.9958 $\pm$ 0.0005 & 0.0873 $\pm$ 0.0051 & 0.9112 $\pm$ 0.0081 & 0.3463 $\pm$ 0.0122 \\
        \cmidrule(lr){3-7}
        & & IBM-Fez & 0.9955 $\pm$ 0.0009 & 0.0875 $\pm$ 0.0136 & 0.9117 $\pm$ 0.0074 & 0.3431 $\pm$ 0.0147 \\
        & & IBM-Marrakesh & 0.9952 $\pm$ 0.0008 & 0.0934 $\pm$ 0.0128 & 0.9117 $\pm$ 0.0067 & 0.3464 $\pm$ 0.0146 \\
        & & IBM-Torino & 0.9952 $\pm$ 0.0010 & 0.0922 $\pm$ 0.0149 & 0.9117 $\pm$ 0.0074 & 0.3457 $\pm$ 0.0074 \\
        & & IBM-Yonsei & 0.9956 $\pm$ 0.0006 & 0.0885 $\pm$ 0.0092 & 0.9112 $\pm$ 0.0082 & 0.3477 $\pm$ 0.0138 \\
        & & IBM-Brisbane & 0.9954 $\pm$ 0.0012 & 0.0892 $\pm$ 0.0177 & 0.9116 $\pm$ 0.0073 & 0.3445 $\pm$ 0.0226 \\
        & & IBM-Brussels & 0.9954 $\pm$ 0.0008 & 0.0898 $\pm$ 0.0103 & 0.9107 $\pm$ 0.0081 & 0.3474 $\pm$ 0.0127 \\
        & & IBM-Strasbourg & 0.9957 $\pm$ 0.0010 & 0.0857 $\pm$ 0.0158 & 0.9106 $\pm$ 0.0072 & 0.3458 $\pm$ 0.0223 \\
        \bottomrule
    \end{tabular}
\end{table}

\subsection{Performance Ranking of Classical NNs and HQNNs in Predicting Key Physicochemical Properties}

To further examine the effects of different learning strategies, we assessed overall rankings of R\textsuperscript{2 }and MAE scores across all classical NNs and HQNNs. (Figure~\ref{fig:fig3_boxplot_ranking}) Firstly, both classical NNs exhibited relatively poor performance, with MLP ranking 6th in both R\textsuperscript{2 }and MAE, while GNN ranked 7th and 8th, respectively. Among HQNN variants, HQSc/4Q performed the worst, ranking 8th in R\textsuperscript{2 }and 4th in MAE while HQFr/4Q showed similar performance, ranking 4th in R\textsuperscript{2 }and 7th in MAE. Among the 4-qubit HQNN variants, HQFi/4Q performed the best, with 3rd in both R\textsuperscript{2 }and MAE. Notably, the 9-qubit HQNN models leveraging pre-trained weights – HQFi/9Q and HQFr/9Q – achieved the highest performance: HQFi/9Q ranked 2nd in R\textsuperscript{2 }and 1st in MAE, whereas HQFr/9Q ranked 1st in R\textsuperscript{2 }and 2nd MAE. In contrast, HQSc/9Q, which also utilized 9 qubits but was trained from scratch, ranked 5th in both R\textsuperscript{2 }and MAE. Overall, these outcomes highlight three key observations. First, in most cases, integrating a quantum layer delivers better performance than the best-performing classical MLP or GNN models. Second, leveraging pre-trained classical weights offers clear advantages over training from scratch. Finally, employing a higher qubit count further enhances predictive accuracy.

\subsection{Impact of Quantum Hardware Noise on HQNN Performance}

To assess the robustness of HQNNs under real quantum hardware conditions, we trained models using various noise configurations from IBM quantum hardware (Figure S1). The noise parameters were extracted from seven IBM quantum hardware, each with a qubit count exceeding 100 and a CLOPS value of at least 150K. These noise settings were utilized to model depolarizing channels, amplitude damping, phase damping, and readout errors, as detailed in Table S8. However, only 4Q models, including HQSc, HQFi, and HQFr, were trained under IBM hardware noise due to hardware constraints. The basicity property was selected for evaluation across all seven noise configurations, as it consistently demonstrated a quantum advantage in both 4Q and 9Q HQNNs (Figure~\ref{fig:fig2_hqnn_performance}). All models for the other four properties were trained using noise from IBM-Fez.

For basicity prediction, 4Q models (HQSc, HQFi, and HQFr) maintained performance comparable to their noiseless counterparts. Notably, HQFi/4Q exhibited improved robustness under various noise settings, with R\textsuperscript{2} gains ranging from 0.26$\%$ to 0.39$\%$ and MAE reductions between 7.18$\%$ and 8.41$\%$ (Table~\ref{tab:noise_pka}). Both HQSc and HQFr also demonstrated resilience in noisy conditions, with HQSc showing improved R\textsuperscript{2} while HQFr maintained its value. However, neither model showed improvements in MAE. Performance under noise conditions varied across different property predictions. In viscosity predictions with HQMLP, all 3 variants exhibited no significant changes in both R\textsuperscript{2} and MAE under the noise. In contrast, noisy HQGNN demonstrated marginal R\textsuperscript{2} improvements, with HQFi achieving the highest increase of 0.86$\%$, alongside a maximum MAE improvement of 0.51$\%$. For vapor pressure, all HQMLP models showed a small decrease in R\textsuperscript{2}. HQFr showed a marginal increase in MAE, whereas HQSc and HQFi demonstrated lower MAE values. For boiling point, HQSc and HQFr showed small improvements in both R\textsuperscript{2 }and MAE values while HQFi was worse than the noiseless counterpart in both metrics. In melting point predictions, HQSc and HQFr exhibited a slight decline in R\textsuperscript{2} while demonstrating improvements in MAE. Conversely, the HQFi model showed an enhancement in R\textsuperscript{2 }but a higher MAE value. While the noise simulations of the HQNN models yielded varying results, the observed changes, both improvements and declines, were marginal and statistically negligible. These findings indicate that the developed VQR method exhibited robustness to hardware-induced noise in simulated environments. 

\section{Discussion}

Quantum advantages have been demonstrated theoretically in areas such as machine learning on quantum data, quantum simulations in natural sciences (e.g., quantum chemistry), and cryptographic code-breaking. However, current quantum computers are hindered by noise, which degrades computational reliability with increasing circuit depth. Many quantum algorithms require fault-tolerant quantum computers, while NISQ-era error mitigation techniques introduce exponential computational complexity, limiting scalability. Despite these challenges, medium-scale systems may still benefit from quantum approaches. Given the heuristic nature of near-term quantum algorithms, rigorous proofs of quantum advantage remain difficult, necessitating experimental validation. Here, we developed HQNNs that integrate quantum circuits with classical architecture. By leveraging classical NN for feature processing and quantum operations for final output generation, our approach reduces the need for deep quantum circuits on NISQ devices. Although noise experiments were used instead of real hardware, results demonstrate the potential of HQMLP and HQGNN architecture. Our findings would bridge the gap between theoretical quantum advantage and practical implementation, underscoring the need for systematic experimental validation and scalable quantum approaches.

Our results showed performance improvements in QSPR modeling for amine solvents, but there are inherent limitations, leaving room for further improvements. Several key challenges should be addressed to fully harness quantum-enhancing modeling. First, our HQNN architecture was built within a limited hyperparameter space. Expanding hyperparameter exploration, particularly in entanglement structures, gate types, and initialization strategies, could further enhance HQNN performance. Additionally, in terms of Ansatz design, adaptive quantum circuits that dynamically adjust to task requirements may improve both performance and generalization. Second, while HQNNs leverage quantum phenomena such as superposition, entanglement, and interference to model complex correlations, the specific quantum mechanisms driving these advantages remain unclear. Performance gains were inconsistent, with a maximum MAE improvement of 7.55$\%$ for basicity prediction, suggesting that iterative QNN architectures alone are insufficient for broader practical adoption. Third, scalability remains a major challenge. Although the iterative QNN framework benefits from a fixed qubit count, it requires iterative training for each sample, which may become infeasible for larger datasets. Most quantum studies focus on small-scale datasets \cite{ref59_bowles2024_arxiv,ref60_mari2020_quantum}, whereas our study addresses real-world applications with an average sample size exceeding 10,000. Without advancements in physical qubit capabilities, training time scales exponentially in both quantum simulations and hardware, making real-world deployment impractical. Exploring alternative architectures, such as variational quantum circuits with parallelized layers, could improve scalability. Fourth, noisy simulations do not fully reflect real hardware conditions. Collaborating with hardware developers to access real-time noise profiles and integrating advanced error mitigation techniques, including zero-noise extrapolation and probabilistic error cancellation, are essential for realistic performance evaluation. Finally, understanding the interplay between classical preprocessing and quantum circuit configurations is critical for efficient HQNN design. Expanding applications beyond amine solvents will test their generalizability in material discovery. As quantum hardware matures, the feasibility of HQNNs in materials science is expected to improve, paving the way for more robust quantum-enhanced modeling techniques.

This study contributes to the growing body of research on QML and QNN by demonstrating both the practical implementation and limitations of HQNNs in QSPR modeling. While our findings do not establish a definitive quantum advantage, they highlight the potential of hybrid quantum-classical architectures for real-world applications. By integrating quantum circuits with classical neural networks, our approach reduces the need for deep quantum circuits on NISQ devices, making quantum-enhanced learning more feasible. However, scalability remains a major challenge, particularly concerning trainability and prediction error bounds. Overcoming these limitations through systematic experimental validation and scalable quantum approaches will be crucial for closing the gap between theoretical potential and practical utility in quantum-enhanced machine learning.
\section{Author Contributions}
H.C., J.K., and H.L. contributed to methodology, validation, and writing. H.L. was responsible for data curation and conceptualization. All authors reviewed the manuscript.

\section{Acknowledgments}

This research was supported by Quantum Advantage challenge research based on Quantum Computing through the National Research Foundation of Korea (NRF) funded by the Ministry of Science and ICT (RS-2023-00257288). H.C., J.K., and H.L. express their gratitude to Professor Kyoung Tai No for his support and guidance.

\section{Declarations}

\textbf{Competing interests \\ }The authors declare no competing interest.

\textbf{Funding}\\
H.C., J.K., and H.L. are supported by Quantum Advantage challenge research based on Quantum Computing through the National Research Foundation of Korea (NRF) funded by the Ministry of Science and ICT (RS-2023-00257288).

\textbf{Code Availability \\ }All results in this work can be found at \href{https://github.com/hclim0213/HQNN-Amine}{https://github.com/hclim0213/HQNN-Amine}.

% \bibliographystyle{unsrt}
% \bibliography{references.bib}

\begin{thebibliography}{10}

\bibitem{ref01_ang2016_energypolicy}
Beng~Wah Ang and Bin Su.
\newblock Carbon emission intensity in electricity production: {A} global analysis.
\newblock {\em Energy Policy}, 94:56--63, 2016.

\bibitem{ref02_bui2018_energyenvsci}
Mai Bui, Claire~S Adjiman, Andr{\'e} Bardow, Edward~J Anthony, Andy Boston, Solomon Brown, Paul~S Fennell, Sabine Fuss, Amparo Galindo, Leigh~A Hackett, et~al.
\newblock Carbon capture and storage ({CCS}): the way forward.
\newblock {\em Energy \& Environmental Science}, 11(5):1062--1176, 2018.

\bibitem{ref03_rochelle2009_science}
Gary~T Rochelle.
\newblock Amine scrubbing for {CO$_{2}$} capture.
\newblock {\em Science}, 325(5948):1652--1654, 2009.

\bibitem{ref04_booras1991_energy}
GS~Booras and SC~Smelser.
\newblock An engineering and economic evaluation of {CO$_{2}$} removal from fossil-fuel-fired power plants.
\newblock {\em Energy}, 16(11-12):1295--1305, 1991.

\bibitem{ref05_porcheron2011_energyprocedia}
Fabien Porcheron, Alexandre Gibert, Marc Jacquin, Pascal Mougin, Abdelaziz Faraj, Aur{\'e}lie Goulon, Pierre-Antoine Bouillon, Bruno Delfort, Dominique Le~Pennec, and Ludovic Raynal.
\newblock {High Throughput Screening} of amine thermodynamic properties applied to post-combustion {CO$_{2}$} capture process evaluation.
\newblock {\em Energy Procedia}, 4:15--22, 2011.

\bibitem{ref06_chowdhury2009_energyprocedia}
Firoz~Alam Chowdhury, Hiromichi Okabe, Shinkichi Shimizu, Masami Onoda, and Yuichi Fujioka.
\newblock Development of novel tertiary amine absorbents for {CO$_{2}$} capture.
\newblock {\em Energy Procedia}, 1(1):1241--1248, 2009.

\bibitem{ref07_chowdhury2011_energyprocedia}
Firoz~Alam Chowdhury, Hiromichi Okabe, Hidetaka Yamada, Masami Onoda, and Yuichi Fujioka.
\newblock Synthesis and selection of hindered new amine absorbents for {CO$_{2}$} capture.
\newblock {\em Energy Procedia}, 4:201--208, 2011.

\bibitem{ref08_gorji2022_chemengresdesign}
Zahra~Eshaghi Gorji, Ali~Ebrahimpoor Gorji, and Siavash Riahi.
\newblock A structure-property model for the prediction of {pKa} values of different amines in the {CO$_{2}$} capture process of concern to the prediction of thermodynamic properties.
\newblock {\em Chemical Engineering Research and Design}, 180:13--24, 2022.

\bibitem{ref09_porcheron2013_ogst}
Fabien Porcheron, Marc Jacquin, Nabil El~Hadri, DA~Saldana, Aur{\'e}lie Goulon, and Abdelaziz Faraj.
\newblock Graph machine based-{QSAR} approach for modeling thermodynamic properties of amines: application to {CO$_{2}$} capture in postcombustion.
\newblock {\em Oil \& Gas Science and Technology--Revue d’IFP Energies nouvelles}, 68(3):469--486, 2013.

\bibitem{ref10_khaheshi2019_indengchemres}
Shima Khaheshi, Siavash Riahi, Mohammad Mohammadi-Khanaposhtani, and Hoda Shokrollahzadeh.
\newblock Prediction of amines capacity for carbon dioxide absorption based on structural characteristics.
\newblock {\em Industrial \& Engineering Chemistry Research}, 58(20):8763--8771, 2019.

\bibitem{ref11_kuenemann2017_molinf}
Melaine~A. Kuenemann and Denis Fourches.
\newblock Cheminformatics modeling of amine solutions for assessing their {CO$_{2}$} absorption properties.
\newblock {\em Molecular Informatics}, 36(7):1600143, 2017.

\bibitem{ref12_biamonte2017_nature}
Jacob Biamonte, Peter Wittek, Nicola Pancotti, Patrick Rebentrost, Nathan Wiebe, and Seth Lloyd.
\newblock Quantum machine learning.
\newblock {\em Nature}, 549(7671):195--202, 2017.

\bibitem{ref13_mcardle2020_revmodphys}
Sam McArdle, Suguru Endo, Al{\'a}n Aspuru-Guzik, Simon~C Benjamin, and Xiao Yuan.
\newblock Quantum computational chemistry.
\newblock {\em Reviews of Modern Physics}, 92(1):015003, 2020.

\bibitem{ref14_nannicini2019_physicalreviewe}
Giacomo Nannicini.
\newblock Performance of hybrid quantum-classical variational heuristics for combinatorial optimization.
\newblock {\em Physical Review E}, 99(1):013304, 2019.

\bibitem{ref15_hirai2024_scientificreports}
Hirotoshi Hirai.
\newblock Practical application of quantum neural network to materials informatics.
\newblock {\em Scientific Reports}, 14(1):8583, 2024.

\bibitem{ref16_ray2024_qce}
Anupama Ray, Dhiraj Madan, Srushti Patil, Pushpak Pati, Marianna Rapsomaniki, Aviwe Kohlakala, Thembelihle~Rose Dlamini, Stephanie~Julia Muller, Kahn Rhrissorrakrai, Filippo Utro, et~al.
\newblock Hybrid quantum-classical graph neural networks for tumor classification in digital pathology.
\newblock In {\em 2024 IEEE International Conference on Quantum Computing and Engineering (QCE)}. IEEE, 2024.

\bibitem{ref17_hirai2023_arxiv}
Hirotoshi Hirai.
\newblock Application of quantum neural network model to a multivariate regression problem.
\newblock {\em arXiv preprint}, 2023.

\bibitem{ref18_beer2020_naturecommunications}
Kerstin Beer, Dmytro Bondarenko, Terry Farrelly, Tobias~J Osborne, Robert Salzmann, Daniel Scheiermann, and Ramona Wolf.
\newblock Training deep quantum neural networks.
\newblock {\em Nature Communications}, 11(1):808, 2020.

\bibitem{ref19_farhi2018_arxiv}
Edward Farhi and Hartmut Neven.
\newblock Classification with quantum neural networks on near term processors.
\newblock {\em arXiv preprint}, 2018.

\bibitem{ref20_cong2019_naturephysics}
Iris Cong, Soonwon Choi, and Mikhail~D Lukin.
\newblock Quantum convolutional neural networks.
\newblock {\em Nature Physics}, 15(12):1273--1278, 2019.

\bibitem{ref21_tomar2025_arxiv}
Sahil Tomar, Rajeshwar Tripathi, and Sandeep Kumar.
\newblock {Comprehensive Survey of QML: From Data Analysis to Algorithmic Advancements}.
\newblock {\em arXiv preprint}, 2025.

\bibitem{ref22_arute2019_nature}
Frank Arute, Kunal Arya, Ryan Babbush, Dave Bacon, Joseph~C Bardin, Rami Barends, Rupak Biswas, Sergio Boixo, Fernando~GSL Brandao, David~A Buell, et~al.
\newblock Quantum supremacy using a programmable superconducting processor.
\newblock {\em Nature}, 574(7779):505--510, 2019.

\bibitem{ref23_zhong2020_science}
Han-Sen Zhong, Hui Wang, Yu-Hao Deng, Ming-Cheng Chen, Li-Chao Peng, Yi-Han Luo, Jian Qin, Dian Wu, Xing Ding, Yi~Hu, et~al.
\newblock Quantum computational advantage using photons.
\newblock {\em Science}, 370(6523):1460--1463, 2020.

\bibitem{ref24_montanaro2016_npjquantuminf}
Ashley Montanaro.
\newblock Quantum algorithms: an overview.
\newblock {\em npj Quantum Information}, 2(1):1--8, 2016.

\bibitem{ref25_yu2010_jcheminf}
Haiying Yu, Ralph K\"{u}hne, Ralf-Uwe Ebert, and Gerrit Sch\"{u}\"{u}rmann.
\newblock Comparative analysis of {QSAR} models for predicting {pKa} of organic oxygen acids and nitrogen bases from molecular structure.
\newblock {\em Journal of Chemical Information and Modeling}, 50(11):1949--1960, 2010.

\bibitem{ref26_baltruschat2020_f1000research}
Marcel Baltruschat and Paul Czodrowski.
\newblock Machine learning meets {pKa}.
\newblock {\em F1000Research}, 9, 2020.

\bibitem{ref27_pan2021_jcheminf}
Xiaolin Pan, Hao Wang, Cuiyu Li, John~ZH Zhang, and Changge Ji.
\newblock {MolGpka}: {A} web server for small molecule {pKa} prediction using a graph-convolutional neural network.
\newblock {\em Journal of Chemical Information and Modeling}, 61(7):3159--3165, 2021.

\bibitem{ref28_mansouri2019_jcheminf}
Kamel Mansouri, Neal~F Cariello, Alexandru Korotcov, Valery Tkachenko, Chris~M Grulke, Catherine~S Sprankle, David Allen, Warren~M Casey, Nicole~C Kleinstreuer, and Antony~J Williams.
\newblock Open-source {QSAR} models for {pKa} prediction using multiple machine learning approaches.
\newblock {\em Journal of Cheminformatics}, 11:1--20, 2019.

\bibitem{ref29_chew2024_jcheminf}
Alex~K Chew, Matthew Sender, Zachary Kaplan, Anand Chandrasekaran, Jackson Chief~Elk, Andrea~R Browning, H~Shaun Kwak, Mathew~D Halls, and Mohammad Atif~Faiz Afzal.
\newblock Advancing material property prediction: using physics-informed machine learning models for viscosity.
\newblock {\em Journal of Cheminformatics}, 16(1):31, 2024.

\bibitem{ref30_epa2012_epa}
U.S. Environmental~Protection Agency.
\newblock Estimation programs interface suite™ for {Microsoft® Windows}, v 4.11.
\newblock United States Environmental Protection Agency, Washington, DC, USA, 2012.

\bibitem{ref31_kazakov2012_ijthermophys}
A~Kazakov, Chris~D Muzny, K~Kroenlein, Vladimir Diky, Robert~D Chirico, Joe~W Magee, Ilmutdin~M Abdulagatov, and M~Frenkel.
\newblock {NIST/TRC} source data archival system: The next-generation data model for storage of thermophysical properties.
\newblock {\em International Journal of Thermophysics}, 33:22--33, 2012.

\bibitem{ref32_williams2017_jcheminf}
Antony~J Williams, Christopher~M Grulke, Jeff Edwards, Andrew~D McEachran, Kamel Mansouri, Nancy~C Baker, Grace Patlewicz, Imran Shah, John~F Wambaugh, Richard~S Judson, et~al.
\newblock The {CompTox Chemistry Dashboard}: a community data resource for environmental chemistry.
\newblock {\em Journal of Cheminformatics}, 9:1--27, 2017.

\bibitem{ref33_landrum2013_rdkit}
Greg Landrum.
\newblock {RDKit: A Software Suite for Cheminformatics, Computational Chemistry, and Predictive Modeling}, 2013.
\newblock Accessed: 2025-02-28.

\bibitem{ref34_durant2002_jcheminf}
Joseph~L Durant, Burton~A Leland, Douglas~R Henry, and James~G Nourse.
\newblock Reoptimization of {MDL} keys for use in drug discovery.
\newblock {\em Journal of Chemical Information and Computer Sciences}, 42(6):1273--1280, 2002.

\bibitem{ref35_gedeck2006_jcheminf}
Peter Gedeck, Bernhard Rohde, and Christian Bartels.
\newblock {QSAR}--how good is it in practice? comparison of descriptor sets on an unbiased cross section of corporate data sets.
\newblock {\em Journal of Chemical Information and Modeling}, 46(5):1924--1936, 2006.

\bibitem{ref36_rogers2010_jcheminf}
David Rogers and Mathew Hahn.
\newblock Extended-connectivity fingerprints.
\newblock {\em Journal of Chemical Information and Modeling}, 50(5):742--754, 2010.

\bibitem{ref37_bolton2008_annrepcompchem}
Evan~E Bolton, Yanli Wang, Paul~A Thiessen, and Stephen~H Bryant.
\newblock {PubChem}: integrated platform of small molecules and biological activities.
\newblock In {\em Annual Reports in Computational Chemistry}, pages 217--241. Elsevier, 2008.

\bibitem{ref38_morgan1965_jchemdoc}
Harry~L. Morgan.
\newblock The generation of a unique machine description for chemical structures-a technique developed at chemical abstracts service.
\newblock {\em Journal of Chemical Documentation}, 5(2):107--113, 1965.

\bibitem{ref39_tailor2021_arxiv}
Shyam~A Tailor, Felix~L Opolka, Pietro Lio, and Nicholas~D Lane.
\newblock Do we need anisotropic graph neural networks?
\newblock {\em arXiv preprint}, 2021.

\bibitem{ref40_carhart1985_jcheminf}
Raymond~E Carhart, Dennis~H Smith, and Rengachari Venkataraghavan.
\newblock Atom pairs as molecular features in structure-activity studies: definition and applications.
\newblock {\em Journal of Chemical Information and Computer Sciences}, 25(2):64--73, 1985.

\bibitem{ref41_goodfellow2016_deeplearning}
Ian Goodfellow.
\newblock Deep learning.
\newblock {\em MIT Press}, 2016.

\bibitem{ref42_agarap2018_arxiv}
Abien~Fred Agarap.
\newblock Deep learning using rectified linear units {(ReLU)}.
\newblock {\em arXiv preprint}, 2018.

\bibitem{ref43_kingma2014_arxiv}
Diederik~P Kingma and Jimmy Ba.
\newblock {Adam}: {A} method for stochastic optimization.
\newblock {\em arXiv preprint}, 2014.

\bibitem{ref44_dai2016_icml}
Hanjun Dai, Bo~Dai, and Le~Song.
\newblock Discriminative embeddings of latent variable models for structured data.
\newblock In {\em International Conference on Machine Learning}. PMLR, 2016.

\bibitem{ref45_brody2021_arxiv}
Shaked Brody, Uri Alon, and Eran Yahav.
\newblock How attentive are graph attention networks?
\newblock {\em arXiv preprint}, 2021.

\bibitem{ref46_chen2020_icml}
Ming Chen, Zhewei Wei, Zengfeng Huang, Bolin Ding, and Yaliang Li.
\newblock Simple and deep graph convolutional networks.
\newblock In {\em International Conference on Machine Learning}. PMLR, 2020.

\bibitem{ref47_shi2020_arxiv}
Yunsheng Shi, Zhengjie Huang, Shikun Feng, Hui Zhong, Wenjin Wang, and Yu~Sun.
\newblock Masked label prediction: {Unified} message passing model for semi-supervised classification.
\newblock {\em arXiv preprint}, 2020.

\bibitem{ref48_nielsen2010_quantumcomp}
Michael~A Nielsen and Isaac~L Chuang.
\newblock {\em Quantum computation and quantum information}.
\newblock Cambridge University Press, 2010.

\bibitem{ref49_bergholm2018_arxiv}
Ville Bergholm, Josh Izaac, Maria Schuld, Christian Gogolin, Shahnawaz Ahmed, Vishnu Ajith, M~Sohaib Alam, Guillermo Alonso-Linaje, B~AkashNarayanan, Ali Asadi, et~al.
\newblock {Pennylane}: Automatic differentiation of hybrid quantum-classical computations.
\newblock {\em arXiv preprint}, 2018.

\bibitem{ref50_zaman2024_arxiv}
Kamila Zaman, Tasnim Ahmed, Muhammad~Abdullah Hanif, Alberto Marchisio, and Muhammad Shafique.
\newblock A comparative analysis of {Hybrid-Quantum Classical Neural Networks}.
\newblock {\em arXiv preprint}, 2024.

\bibitem{ref51_rezaei2016_jngse}
Bijan Rezaei and Siavash Riahi.
\newblock Prediction of {CO$_{2}$} loading of amines in carbon capture process using membrane contactors: {A} molecular modeling.
\newblock {\em Journal of Natural Gas Science and Engineering}, 33:388--396, 2016.

\bibitem{ref52_bernhardsen2017_ijggc}
Ida~M Bernhardsen and Hanna~K Knuutila.
\newblock A review of potential amine solvents for {CO$_{2}$} absorption process: {Absorption} capacity, cyclic capacity and {pKa}.
\newblock {\em International Journal of Greenhouse Gas Control}, 61:27--48, 2017.

\bibitem{ref53_yuan2017_energyprocedia}
Ye~Yuan, Brent Sherman, and Gary~T Rochelle.
\newblock Effects of viscosity on {CO$_{2}$} absorption in aqueous piperazine/2-methylpiperazine.
\newblock {\em Energy Procedia}, 114:2103--2120, 2017.

\bibitem{ref54_li2014_ijggc}
Han Li, Yann Le~Moullec, Jiahui Lu, Jian Chen, Jose Carlos~Valle Marcos, and Guofei Chen.
\newblock Solubility and energy analysis for {CO$_{2}$} absorption in piperazine derivatives and their mixtures.
\newblock {\em International Journal of Greenhouse Gas Control}, 31:25--32, 2014.

\bibitem{ref55_wu2013_exptthermfluid}
Sheng-Hong Wu, Alvin~R Caparanga, Rhoda~B Leron, and Meng-Hui Li.
\newblock Vapor pressures of aqueous blended-amine solutions containing {(TEA/AMP/MDEA)+(DEA/MEA/PZ) at temperatures (303.15--343.15) K}.
\newblock {\em Experimental Thermal and Fluid Science}, 48:1--7, 2013.

\bibitem{ref56_freeman2010_ijggc}
Stephanie~A Freeman, Ross Dugas, David~H Van~Wagener, Thu Nguyen, and Gary~T Rochelle.
\newblock Carbon dioxide capture with concentrated, aqueous piperazine.
\newblock {\em International Journal of Greenhouse Gas Control}, 4(2):119--124, 2010.

\bibitem{ref57_yang2017_chemrev}
Xin Yang, Robert~J Rees, William Conway, Graeme Puxty, Qi~Yang, and David~A Winkler.
\newblock Computational modeling and simulation of {CO$_{2}$} capture by aqueous amines.
\newblock {\em Chemical Reviews}, 117(14):9524--9593, 2017.

\bibitem{ref58_katritzky2001_crystgrowthdes}
Alan~R Katritzky, Ritu Jain, Andre Lomaka, Ruslan Petrukhin, Uko Maran, and Mati Karelson.
\newblock Perspective on the relationship between melting points and chemical structure.
\newblock {\em Crystal Growth \& Design}, 1(4):261--265, 2001.

\bibitem{ref59_bowles2024_arxiv}
Joseph Bowles, Shahnawaz Ahmed, and Maria Schuld.
\newblock Better than classical? the subtle art of benchmarking quantum machine learning models.
\newblock {\em arXiv preprint}, 2024.

\bibitem{ref60_mari2020_quantum}
Andrea Mari, Thomas~R Bromley, Josh Izaac, Maria Schuld, and Nathan Killoran.
\newblock Transfer learning in hybrid classical-quantum neural networks.
\newblock {\em Quantum}, 4:340, 2020.

\end{thebibliography}

%%% Remove comment to use the external .bib file (using bibtex).
%%% and comment out the ``thebibliography'' section.

\appendix
\clearpage
\thispagestyle{empty}
\renewcommand{\thetable}{S\arabic{table}}
\renewcommand{\thefigure}{S\arabic{figure}}
\setcounter{table}{0}
\setcounter{figure}{0}

\section*{Supporting Information for}

\title{Hybrid Quantum Neural Networks with Variational Quantum Regressor for Enhancing QSPR Modeling of CO\textsubscript{2}-Capturing Amine}
\maketitle

\author[1]{\textbf{Hyein Cho}$^{\dagger}$}
\author[2]{\textbf{Jeonghoon Kim}$^{\dagger}$}
\author[2]{\textbf{Hocheol Lim}$^{\ast}$}

\begin{flushleft}
    $^{\dagger}$ These authors contributed equally to this work. \\
    $^{\ast}$ Corresponding author: Hocheol Lim (ihc0213@yonsei.ac.kr)
\end{flushleft}

\vspace{0.5cm}

\textbf{\ \ \ \ }The supporting information for ‘Hybrid Quantum Neural Networks with Variational Quantum Regressor for Enhancing QSPR Modeling of CO\textsubscript{2}-Capturing Amine’ includes Figure S1 for IBM quantum hardware noise configuration applied to hybrid quantum neural network architecture, Table S1 for the hyperparameter tuning procedure, Table S2 for the optimal hyperparameters for classical neural networks, Table S3–S7 for performance metrics for classical neural networks and hybrid quantum neural networks in predicting basicity, viscosity, vapor pressure, boiling point, and melting point of CO\textsubscript{2}-capturing amine, Table S8 for noise parameters derived from IBM quantum hardware, and Table S9 for hybrid quantum neural network performance under noiseless and noisy conditions.

\ \ \ \ In this study, there are many abbreviations as follows. CCS; Carbon Capture and Storage, DMPNN; Directed Message-passing Neural Network, ECFP6; Extended-Connectivity Fingerprints, EGC; Efficient Graph Convolutional Network, FCFP4; Functional-Class Fingerprints, GAT; Graph Attention Transformer, GCN; Graph Convolutional Network, GNN; Graph Neural Network, HQDMPNN; Hybrid Quantum Directed Message-passing Neural Network, HQFi; HQNN Pretrain-Finetuned, HQFr; HQNN Pretrain-Frozen, HQGNN; Hybrid Quantum Graph Neural Network, HQMLP; Hybrid Quantum Multi-Layer Perceptron, HQNN; Hybrid Quantum Neural Network, HQSc; HQNN Scratch, MACCS; Molecular Access System, MAE; Mean Absolute Error, MLP; Multi-Layer Perceptron, NISQ; Noisy Intermediate-Scale Quantum, PCFP; PubChem Fingerprints, QCNN; Quantum Convolutional Neural Network, QML; Quantum Machine Learning, QNN; Quantum Neural Network, QResNet; Quantum ResNet, QSPR; Quantitative structure-property relationship, QuanNN; Quanvolutional Neural Network, TCN; Transformer Convolutional Network, VQC; Variational Quantum Circuit, VQR; Variational Quantum Regressor.

\newpage
\begin{figure}[ht!]
  \centering
  \includegraphics[width=1.0\textwidth]{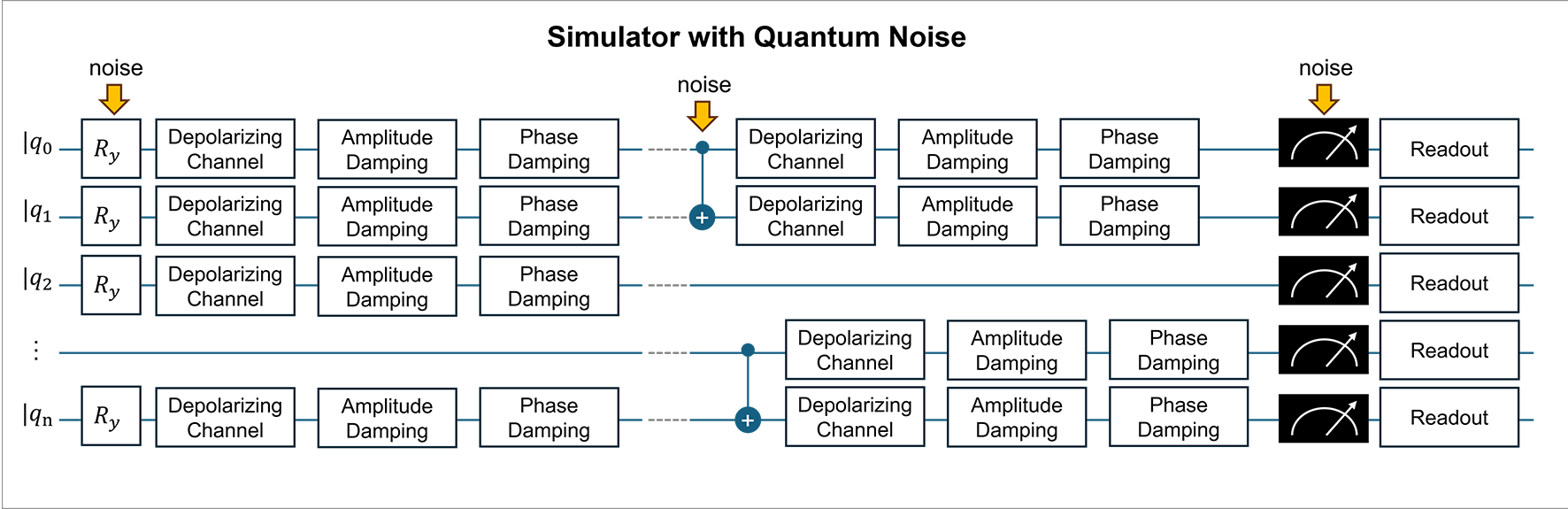}
  \caption{Schematic Representation of IBM Quantum Hardware Noise Configuration in HQNN Models}
\end{figure}

\newpage
\begin{table}
    \centering
    \caption{Hyperparameters Used in Hyperparameter Tuning Procedure}
    \setlength{\tabcolsep}{30pt}
    \renewcommand{\arraystretch}{1.3}
    \begin{tabular}{ccc}
        \toprule
        \textbf{Model} & \textbf{Tuning Parameters} & \textbf{Fixed Parameters} \\
        \specialrule{1.5pt}{0pt}{0pt}
        \multirow{8}{*}{MLP} & hidden layers = & \multirow{8}{*}{\shortstack{batch size = 128\\learning rate = 0.001\\epochs = 300}} \\
        & [2048, 1024, 512, 256, 128], & \\
        & [2048, 1024, 512, 256], & \\
        & [2048, 1024, 512], & \\
        & [1024, 1024, 1024], & \\
        & [1024, 512, 256, 128], & \\
        & [1024, 512, 256] & \\
        & dropout rate = 0.1, 0.2 & \\
        \cmidrule(lr){1-3}
        \multirow{4}{*}{DMPNN} & \multirow{4}{*}{\shortstack{hidden layers = 512, 1024, 2048\\step = 1, 2, 4}} & dropout rate = 0.1 \\
        & & batch size = 128 \\
        & & learning rate = 0.001 \\
        & & learning rate = 0.001 \\
        \cmidrule(lr){1-3}
        \multirow{4}{*}{EGC} & \multirow{4}{*}{\shortstack{hidden layers = 512, 1024, 2048\\graph layers = 1, 2, 3, 4\\graph heads = 1, 2, 4, 8}} & dropout rate = 0.1 \\
        & & batch size = 128 \\
        & & learning rate = 0.001 \\
        & & learning rate = 0.001 \\
        \cmidrule(lr){1-3}
        \multirow{4}{*}{GAT} & \multirow{4}{*}{\shortstack{hidden layers = 512, 1024, 2048\\graph layers = 1, 2, 3, 4\\graph heads = 1, 2, 4, 8}} & dropout rate = 0.1 \\
        & & batch size = 128 \\
        & & learning rate = 0.001 \\
        & & learning rate = 0.001 \\
        \cmidrule(lr){1-3}
        \multirow{4}{*}{GCN} & \multirow{4}{*}{\shortstack{hidden layers = 512, 1024, 2048\\graph layers = 1, 2, 3, 4\\graph heads = 1, 2, 4, 8}} & dropout rate = 0.1 \\
        & & batch size = 128 \\
        & & learning rate = 0.001 \\
        & & learning rate = 0.001 \\
        \cmidrule(lr){1-3}
        \multirow{4}{*}{TCN} & \multirow{4}{*}{\shortstack{hidden layers = 512, 1024, 2048\\graph layers = 1, 2, 3, 4\\graph heads = 1, 2, 4, 8}} & dropout rate = 0.1 \\
        & & batch size = 128 \\
        & & learning rate = 0.001 \\
        & & learning rate = 0.001 \\
        \bottomrule
    \end{tabular}
\end{table}

\newpage
\begin{table}
    \centering
    \caption{Optimal Hyperparameters of Classical NN for Prediction of Five Properties}
    \setlength{\tabcolsep}{4pt}
    \renewcommand{\arraystretch}{1.3}
    \begin{tabular}{ccc}
        \toprule
        \textbf{Property} & \textbf{Model} & \textbf{Optimal Parameter} \\
        \specialrule{1.5pt}{0pt}{0pt}
        \multirow{2}{*}{\shortstack{Basicity\\(pKa)}} & MLP & {'fingerprint': ‘MACCS+Avalon’, 'hidden layers': [1024-512-256], 'dropout rate': 0.1} \\
        \cmidrule(lr){2-3}
        & GCN & {‘hidden layers’: 2048, ‘graph layers’: 4} \\
        \cmidrule(lr){1-3}
        \multirow{2}{*}{Viscosity} & MLP & {'fingerprint': ‘rPair’, 'hidden layers': [1024-512-256-128], 'dropout rate': 0.1} \\
        \cmidrule(lr){2-3}
        & DMPNN & {'hidden layers’: 512, 'step': 1} \\
        \cmidrule(lr){1-3}
        \multirow{2}{*}{\shortstack{Vapor\\Pressure}} & MLP & {'fingerprint': ‘rPair’, 'hidden layers': [1024-512-256-128], 'dropout rate': 0.2} \\
        \cmidrule(lr){2-3}
        & DMPNN & {'hidden layers’: 512, 'step': 1} \\
        \cmidrule(lr){1-3}
        \multirow{2}{*}{\shortstack{Boiling\\Point}} & MLP & {'fingerprint': ‘MACCS+PCFP’, 'hidden layers': [2048-1024-512-256], 'dropout rate': 0.1} \\
        \cmidrule(lr){2-3}
        & EGC & {'hidden layers’: 1024, 'graph layers': 4, 'graph heads': 4} \\
        \cmidrule(lr){1-3}
        \multirow{2}{*}{\shortstack{Melting\\Point}} & MLP & {'fingerprint': ‘MACCS+Avalon’, 'hidden layers': [2048-1024-512-256-128], 'dropout rate': 0.1} \\
        \cmidrule(lr){2-3}
        & EGC & {'hidden layers’: 1024, 'graph layers': 2, 'graph heads': 8} \\
        \bottomrule
    \end{tabular}
\end{table}

\newpage
\begin{sidewaystable}
    \centering
    \caption{Prediction Performance of Classical NN and HQNN Models for Basicity (pKa)}
    \vspace{5pt}
    \setlength{\tabcolsep}{4pt}
    \renewcommand{\arraystretch}{1.3}
    \begin{tabular}{ccccccc}
        \toprule
        \textbf{Task} & \textbf{Type} & \textbf{Model} & \textbf{Train R$^2$} & \textbf{Train MAE} & \textbf{Test R$^2$} & \textbf{Test MAE} \\
        \specialrule{1.5pt}{0pt}{0pt}
        \multirow{9}{*}{\shortstack{Basicity\\(pKa)}}
        & \multirow{3}{*}{Classical NN} 
        & MLP & 0.9929 $\pm$ 0.0012 & 0.1261 $\pm$ 0.0127 & 0.9082 $\pm$ 0.0083 & 0.3746 $\pm$ 0.0117 \\
        & & GCN & 0.9908 $\pm$ 0.0010 & 0.1458 $\pm$ 0.0103 & 0.8912 $\pm$ 0.0073 & 0.4263 $\pm$ 0.0109 \\
        & & MLP + GCN & 0.9511 $\pm$ 0.0042 & 0.3806 $\pm$ 0.0207 & 0.8546 $\pm$ 0.0090 & 0.5707 $\pm$ 0.0156 \\
        \cmidrule(lr){2-7}
        & \multirow{6}{*}{\shortstack{HQNN\\(HQMLP)}} 
        & HQSc/4Q & 0.9958 $\pm$ 0.0004 & 0.0870 $\pm$ 0.0070 & 0.9088 $\pm$ 0.0057 & 0.3514 $\pm$ 0.0107 \\
        & & HQFi/4Q & 0.9958 $\pm$ 0.0005 & 0.0873 $\pm$ 0.0051 & 0.9112 $\pm$ 0.0081 & 0.3463 $\pm$ 0.0122 \\
        & & HQFr/4Q & 0.9939 $\pm$ 0.0007 & 0.1141 $\pm$ 0.0098 & 0.9093 $\pm$ 0.0086 & 0.3663 $\pm$ 0.0141 \\
        & & HQSc/9Q & 0.9951 $\pm$ 0.0007 & 0.0965 $\pm$ 0.0116 & 0.9079 $\pm$ 0.0105 & 0.3583 $\pm$ 0.0197 \\
        & & HQFi/9Q & 0.9957 $\pm$ 0.0009 & 0.0885 $\pm$ 0.0128 & 0.9094 $\pm$ 0.0095 & 0.3510 $\pm$ 0.0181 \\
        & & HQFr/9Q & 0.9938 $\pm$ 0.0008 & 0.1155 $\pm$ 0.0105 & 0.9093 $\pm$ 0.0086 & 0.3665 $\pm$ 0.0137 \\
        \bottomrule
    \end{tabular}
\end{sidewaystable}

\newpage
\begin{sidewaystable}
    \centering
    \caption{Prediction Performance of Classical NN and HQNN Models for Viscosity}
    \vspace{5pt}
    \setlength{\tabcolsep}{4pt}
    \renewcommand{\arraystretch}{1.3}
    \begin{tabular}{ccccccc}
        \toprule
        \textbf{Task} & \textbf{Type} & \textbf{Model} & \textbf{Train R$^2$} & \textbf{Train MAE} & \textbf{Test R$^2$} & \textbf{Test MAE} \\
        \specialrule{1.5pt}{0pt}{0pt}
        \multirow{15}{*}{Viscosity}
        & \multirow{3}{*}{Classical NN} 
        & MLP & 0.8282 $\pm$ 0.0024 & 0.1330 $\pm$ 0.0018 & 0.7247 $\pm$ 0.0142 & 0.1771 $\pm$ 0.0067 \\
        & & DMPNN & 0.7967 $\pm$ 0.0037 & 0.1520 $\pm$ 0.0029 & 0.7304 $\pm$ 0.0142 & 0.1786 $\pm$ 0.0057 \\
        & & MLP + DMPNN & 0.8188 $\pm$ 0.0023 & 0.1409 $\pm$ 0.0015 & 0.7149 $\pm$ 0.0177 & 0.1798 $\pm$ 0.0078 \\
        \cmidrule(lr){2-7}
        & \multirow{6}{*}{\shortstack{HQNN\\(HQMLP)}} 
        & HQSc/4Q & 0.8260 $\pm$ 0.0058 & 0.1345 $\pm$ 0.0045 & 0.7269 $\pm$ 0.0161 & 0.1773 $\pm$ 0.0070 \\
        & & HQFi/4Q & 0.8289 $\pm$ 0.0012 & 0.1321 $\pm$ 0.0015 & 0.7247 $\pm$ 0.0158 & 0.1765 $\pm$ 0.0066 \\
        & & HQFr/4Q & 0.8286 $\pm$ 0.0022 & 0.1325 $\pm$ 0.0018 & 0.7257 $\pm$ 0.0137 & 0.1767 $\pm$ 0.0064 \\
        & & HQSc/9Q & 0.8262 $\pm$ 0.0044 & 0.1339 $\pm$ 0.0029 & 0.7205 $\pm$ 0.0174 & 0.1783 $\pm$ 0.0075 \\
        & & HQFi/9Q & 0.8281 $\pm$ 0.0021 & 0.1330 $\pm$ 0.0013 & 0.7268 $\pm$ 0.0153 & 0.1765 $\pm$ 0.0067 \\
        & & HQFr/9Q & 0.8286 $\pm$ 0.0021 & 0.1326 $\pm$ 0.0019 & 0.7263 $\pm$ 0.0138 & 0.1767 $\pm$ 0.0064 \\
        \cmidrule(lr){2-7}
        & \multirow{6}{*}{\shortstack{HQNN\\(HQGNN)}} 
        & HQSc/4Q & 0.8057 $\pm$ 0.0101 & 0.1477 $\pm$ 0.0051 & 0.7249 $\pm$ 0.0105 & 0.1777 $\pm$ 0.0040 \\
        & & HQFi/4Q & 0.8139 $\pm$ 0.0049 & 0.1421 $\pm$ 0.0021 & 0.7239 $\pm$ 0.0134 & 0.1775 $\pm$ 0.0072 \\
        & & HQFr/4Q & 0.8135 $\pm$ 0.0040 & 0.1424 $\pm$ 0.0017 & 0.7252 $\pm$ 0.0148 & 0.1783 $\pm$ 0.0071 \\
        & & HQSc/9Q & 0.8126 $\pm$ 0.0050 & 0.1442 $\pm$ 0.0020 & 0.7334 $\pm$ 0.0130 & 0.1758 $\pm$ 0.0062 \\
        & & HQFi/9Q & 0.8184 $\pm$ 0.0034 & 0.1396 $\pm$ 0.0019 & 0.7287 $\pm$ 0.0131 & 0.1768 $\pm$ 0.0061 \\
        & & HQFr/9Q & 0.8173 $\pm$ 0.0027 & 0.1403 $\pm$ 0.0016 & 0.7307 $\pm$ 0.0124 & 0.1759 $\pm$ 0.0058 \\
        \bottomrule
    \end{tabular}
\end{sidewaystable}

\newpage
\begin{sidewaystable}
    \centering
    \caption{Prediction Performance of Classical NN and HQNN Models for Vapor Pressure}
    \vspace{5pt}
    \setlength{\tabcolsep}{4pt}
    \renewcommand{\arraystretch}{1.3}
    \begin{tabular}{ccccccc}
        \toprule
        \textbf{Task} & \textbf{Type} & \textbf{Model} & \textbf{Train R$^2$} & \textbf{Train MAE} & \textbf{Test R$^2$} & \textbf{Test MAE} \\
        \specialrule{1.5pt}{0pt}{0pt}
        \multirow{9}{*}{\shortstack{Vapor\\Pressure}}
        & \multirow{3}{*}{Classical NN} 
        & MLP & 0.9921 $\pm$ 0.0028 & 0.2064 $\pm$ 0.0449 & 0.8860 $\pm$ 0.0246 & 0.6868 $\pm$ 0.0310 \\
        & & DMPNN & 0.9682 $\pm$ 0.0068 & 0.3886 $\pm$ 0.0567 & 0.8842 $\pm$ 0.0153 & 0.8176 $\pm$ 0.0553 \\
        & & MLP + DMPNN & 0.9711 $\pm$ 0.0040 & 0.4599 $\pm$ 0.0444 & 0.8619 $\pm$ 0.0202 & 0.8401 $\pm$ 0.0346 \\
        \cmidrule(lr){2-7}
        & \multirow{6}{*}{\shortstack{HQNN\\(HQMLP)}} 
        & HQSc/4Q & 0.9857 $\pm$ 0.0076 & 0.2270 $\pm$ 0.0521 & 0.8565 $\pm$ 0.0404 & 0.7586 $\pm$ 0.0623 \\
        & & HQFi/4Q & 0.9937 $\pm$ 0.0038 & 0.1600 $\pm$ 0.0432 & 0.8838 $\pm$ 0.0262 & 0.6896 $\pm$ 0.0345 \\
        & & HQFr/4Q & 0.9906 $\pm$ 0.0045 & 0.2421 $\pm$ 0.0658 & 0.8824 $\pm$ 0.0253 & 0.7220 $\pm$ 0.0403 \\
        & & HQSc/9Q & 0.9889 $\pm$ 0.0096 & 0.2184 $\pm$ 0.0738 & 0.8757 $\pm$ 0.0285 & 0.7238 $\pm$ 0.0486 \\
        & & HQFi/9Q & 0.9944 $\pm$ 0.0007 & 0.1716 $\pm$ 0.0190 & 0.8821 $\pm$ 0.0281 & 0.6752 $\pm$ 0.0300 \\
        & & HQFr/9Q & 0.9911 $\pm$ 0.0033 & 0.2342 $\pm$ 0.0496 & 0.8829 $\pm$ 0.0247 & 0.7124 $\pm$ 0.0271 \\
        \bottomrule
    \end{tabular}
\end{sidewaystable}

\newpage
\begin{sidewaystable}
    \centering
    \caption{Prediction Performance of Classical NN and HQNN Models for Boiling Point}
    \vspace{5pt}
    \setlength{\tabcolsep}{4pt}
    \renewcommand{\arraystretch}{1.3}
    \begin{tabular}{ccccccc}
        \toprule
        \textbf{Task} & \textbf{Type} & \textbf{Model} & \textbf{Train R$^2$} & \textbf{Train MAE} & \textbf{Test R$^2$} & \textbf{Test MAE} \\
        \specialrule{1.5pt}{0pt}{0pt}
        \multirow{9}{*}{\shortstack{Boiling\\Point}}
        & \multirow{3}{*}{Classical NN} 
        & MLP & 0.9731 $\pm$ 0.0026 & 12.7938 $\pm$ 0.5953 & 0.8754 $\pm$ 0.0290 & 21.1258 $\pm$ 0.7807 \\
        & & EGC & 0.9646 $\pm$ 0.0091 & 12.9562 $\pm$ 1.4271 & 0.8245 $\pm$ 0.0217 & 24.4463 $\pm$ 0.7385 \\
        & & MLP + EGC & 0.9738 $\pm$ 0.0023 & 14.1789 $\pm$ 0.6189 & 0.8359 $\pm$ 0.0664 & 21.3045 $\pm$ 0.4688 \\
        \cmidrule(lr){2-7}
        & \multirow{6}{*}{\shortstack{HQNN\\(HQMLP)}} 
        & HQSc/4Q & 0.9713 $\pm$ 0.0077 & 13.5211 $\pm$ 1.9935 & 0.8737 $\pm$ 0.0330 & 21.4920 $\pm$ 1.0453 \\
        & & HQFi/4Q & 0.9755 $\pm$ 0.0052 & 13.4047 $\pm$ 1.6366 & 0.8740 $\pm$ 0.0190 & 21.9557 $\pm$ 1.7384 \\
        & & HQFr/4Q & 0.9748 $\pm$ 0.0038 & 13.6211 $\pm$ 1.0202 & 0.8770 $\pm$ 0.0279 & 21.8153 $\pm$ 0.6905 \\
        & & HQSc/9Q & 0.9688 $\pm$ 0.0068 & 15.9883 $\pm$ 2.1881 & 0.8788 $\pm$ 0.0257 & 22.9442 $\pm$ 1.1271 \\
        & & HQFi/9Q & 0.9729 $\pm$ 0.0071 & 14.8899 $\pm$ 2.3408 & 0.8711 $\pm$ 0.0300 & 22.9561 $\pm$ 0.7798 \\
        & & HQFr/9Q & 0.9755 $\pm$ 0.0029 & 13.1890 $\pm$ 0.8603 & 0.8772 $\pm$ 0.0274 & 21.4763 $\pm$ 0.7304 \\
        \bottomrule
    \end{tabular}
\end{sidewaystable}

\newpage
\begin{sidewaystable}
    \centering
    \caption{Prediction Performance of Classical NN and HQNN Models for Melting Point}
    \vspace{5pt}
    \setlength{\tabcolsep}{4pt}
    \renewcommand{\arraystretch}{1.3}
    \begin{tabular}{ccccccc}
        \toprule
        \textbf{Task} & \textbf{Type} & \textbf{Model} & \textbf{Train R$^2$} & \textbf{Train MAE} & \textbf{Test R$^2$} & \textbf{Test MAE} \\
        \specialrule{1.5pt}{0pt}{0pt}
        \multirow{9}{*}{\shortstack{Melting\\Point}}
        & \multirow{3}{*}{Classical NN} 
        & MLP & 0.9789 $\pm$ 0.0045 & 9.4397 $\pm$ 0.8163 & 0.7506 $\pm$ 0.0307 & 36.5919 $\pm$ 0.7485 \\
        & & EGC & 0.9383 $\pm$ 0.0048 & 21.0883 $\pm$ 1.0484 & 0.7619 $\pm$ 0.0624 & 35.4447 $\pm$ 1.1359 \\
        & & MLP + EGC & 0.9669 $\pm$ 0.0153 & 15.7532 $\pm$ 4.3265 & 0.7063 $\pm$ 0.0723 & 36.5934 $\pm$ 1.0424 \\
        \cmidrule(lr){2-7}
        & \multirow{6}{*}{\shortstack{HQNN\\(HQGNN)}} 
        & HQSc/4Q & 0.9179 $\pm$ 0.0043 & 25.1170 $\pm$ 1.0025 & 0.7554 $\pm$ 0.0358 & 34.9999 $\pm$ 1.1913 \\
        & & HQFi/4Q & 0.9147 $\pm$ 0.0319 & 23.8698 $\pm$ 3.4422 & 0.7715 $\pm$ 0.0220 & 35.8185 $\pm$ 1.0808 \\
        & & HQFr/4Q & 0.9211 $\pm$ 0.0210 & 23.2359 $\pm$ 2.3165 & 0.7530 $\pm$ 0.0370 & 35.9148 $\pm$ 1.0196 \\
        & & HQSc/9Q & 0.9138 $\pm$ 0.0121 & 25.6608 $\pm$ 1.6739 & 0.7419 $\pm$ 0.0382 & 35.4849 $\pm$ 0.8564 \\
        & & HQFi/9Q & 0.9325 $\pm$ 0.0191 & 21.4733 $\pm$ 2.7960 & 0.7887 $\pm$ 0.0265 & 35.3202 $\pm$ 0.8253 \\
        & & HQFr/9Q & 0.9296 $\pm$ 0.0177 & 22.3340 $\pm$ 2.9460 & 0.7804 $\pm$ 0.0186 & 35.4365 $\pm$ 0.8656 \\
        \bottomrule
    \end{tabular}
\end{sidewaystable}

\newpage
\begin{sidewaystable}
    \centering
    \caption{Noise Configurations of IBM Quantum Hardware}
    \vspace{5pt}
    \setlength{\tabcolsep}{4pt}
    \renewcommand{\arraystretch}{1.3}
    \begin{tabular}{ccccccccc}
        \toprule
        \textbf{\shortstack{Quantum\\Hardware}} & \textbf{Qubit} & \textbf{CLOPS} & \textbf{\shortstack{Two Qubit\\Error Rate}} & \textbf{\shortstack{Median SX\\Error Rate}} & \textbf{\shortstack{Readout\\Error Rate}} & \textbf{\shortstack{Median T1\\(sec)}} & \textbf{\shortstack{Median T2\\(sec)}} & \textbf{\shortstack{Gate Time\\(sec)}} \\
        \specialrule{1.5pt}{0pt}{0pt}
        IBM-Fez & 156 & 195K & 2.792 $\times$ 10\textsuperscript{-3} & 2.703 $\times$ 10\textsuperscript{-4} & 1.645 $\times$ 10\textsuperscript{-2} & 1.181 $\times$ 10\textsuperscript{-4} & 9.141 $\times$ 10\textsuperscript{-5} & 6.800 $\times$ 10\textsuperscript{-8} \\
        IBM-Marrakesh & 156 & 195K & 3.410 $\times$ 10\textsuperscript{-3} & 2.460 $\times$ 10\textsuperscript{-4} & 1.540 $\times$ 10\textsuperscript{-2} & 1.780 $\times$ 10\textsuperscript{-4} & 1.139 $\times$ 10\textsuperscript{-4} & 6.800 $\times$ 10\textsuperscript{-8} \\
        IBM-Torino & 133 & 210K & 6.250 $\times$ 10\textsuperscript{-3} & 3.508 $\times$ 10\textsuperscript{-4} & 2.000 $\times$ 10\textsuperscript{-2} & 1.661 $\times$ 10\textsuperscript{-4} & 1.358 $\times$ 10\textsuperscript{-4} & 6.800 $\times$ 10\textsuperscript{-8} \\
        IBM-Yonsei & 127 & 230K & 3.890 $\times$ 10\textsuperscript{-2} & 2.080 $\times$ 10\textsuperscript{-2} & 2.080 $\times$ 10\textsuperscript{-2} & 2.415 $\times$ 10\textsuperscript{-4} & 1.540 $\times$ 10\textsuperscript{-4} & 8.400 $\times$ 10\textsuperscript{-8} \\
        IBM-Brisbane & 127 & 180K & 1.650 $\times$ 10\textsuperscript{-2} & 2.549 $\times$ 10\textsuperscript{-4} & 1.440 $\times$ 10\textsuperscript{-2} & 2.239 $\times$ 10\textsuperscript{-4} & 1.395 $\times$ 10\textsuperscript{-4} & 6.600 $\times$ 10\textsuperscript{-7} \\
        IBM-Brussels & 127 & 220K & 2.860 $\times$ 10\textsuperscript{-2} & 2.822 $\times$ 10\textsuperscript{-2} & 2.420 $\times$ 10\textsuperscript{-2} & 2.667 $\times$ 10\textsuperscript{-4} & 1.222 $\times$ 10\textsuperscript{-4} & 6.600 $\times$ 10\textsuperscript{-7} \\
        IBM-Strasbourg & 127 & 220K & 2.910 $\times$ 10\textsuperscript{-2} & 2.649 $\times$ 10\textsuperscript{-2} & 1.840 $\times$ 10\textsuperscript{-2} & 2.665 $\times$ 10\textsuperscript{-4} & 1.488 $\times$ 10\textsuperscript{-4} & 6.600 $\times$ 10\textsuperscript{-7} \\

        \bottomrule
    \end{tabular}
\end{sidewaystable}

\newpage
\begin{sidewaystable}
    \centering
    \caption{Prediction Performance of HQNN Models for Physicochemical Properties Across IBM Quantum Hardware Noise}
    \vspace{5pt}
    \setlength{\tabcolsep}{4pt}
    \renewcommand{\arraystretch}{1.3}
    \begin{tabular}{cccccccc}
        \toprule
        \textbf{Task} & \textbf{Model} & \textbf{Noise} & \textbf{Type} & \textbf{Train R$^2$} & \textbf{Train MAE} & \textbf{Test R$^2$} & \textbf{Test MAE} \\
        \specialrule{1.5pt}{0pt}{0pt}
        \multirow{2}{*}{Basicity} & \multirow{2}{*}{HQMLP} & \multirow{2}{*}{IBM-Fez} 
        & HQSc/4Q & 0.9946 $\pm$ 0.0011 & 0.1009 $\pm$ 0.0150 & 0.9111 $\pm$ 0.0075 & 0.3517 $\pm$ 0.0174 \\
        & & & HQFr/4Q & 0.9939 $\pm$ 0.0007 & 0.1140 $\pm$ 0.0098 & 0.9093 $\pm$ 0.0086 & 0.3664 $\pm$ 0.0140 \\
        \cmidrule(lr){1-8}
        \multirow{8}{*}{Viscosity} & \multirow{4}{*}{HQMLP} & \multirow{1}{*}{Noiseless}
        & HQFi/4Q & 0.8289 $\pm$ 0.0012 & 0.1321 $\pm$ 0.0015 & 0.7247 $\pm$ 0.0158 & 0.1765 $\pm$ 0.0066 \\
        \cmidrule(lr){3-8}
        & & \multirow{3}{*}{IBM-Fez}
        & HQSc/4Q & 0.8270 $\pm$ 0.0020 & 0.1341 $\pm$ 0.0015 & 0.7265 $\pm$ 0.0165 & 0.1761 $\pm$ 0.0071 \\
        & & & HQFi/4Q & 0.8294 $\pm$ 0.0021 & 0.1315 $\pm$ 0.0012 & 0.7221 $\pm$ 0.0161 & 0.1773 $\pm$ 0.0067 \\
        & & & HQFr/4Q & 0.8286 $\pm$ 0.0022 & 0.1325 $\pm$ 0.0018 & 0.7258 $\pm$ 0.0139 & 0.1767 $\pm$ 0.0064 \\
        \cmidrule(lr){2-8}
        & \multirow{4}{*}{HQGNN} & \multirow{1}{*}{Noiseless}
        & HQFi/4Q & 0.8139 $\pm$ 0.0049 & 0.1421 $\pm$ 0.0021 & 0.7239 $\pm$ 0.0134 & 0.1775 $\pm$ 0.0072 \\
        \cmidrule(lr){3-8}
        & & \multirow{3}{*}{IBM-Fez}
        & HQSc/4Q & 0.8081 $\pm$ 0.0074 & 0.1464 $\pm$ 0.0043 & 0.7276 $\pm$ 0.0139 & 0.1768 $\pm$ 0.0047 \\
        & & & HQFi/4Q & 0.8059 $\pm$ 0.0049 & 0.1472 $\pm$ 0.0019 & 0.7301 $\pm$ 0.0147 & 0.1773 $\pm$ 0.0072 \\
        & & & HQFr/4Q & 0.8077 $\pm$ 0.0046 & 0.1464 $\pm$ 0.0019 & 0.7311 $\pm$ 0.0114 & 0.1774 $\pm$ 0.0057 \\
        \cmidrule(lr){1-8}
        \multirow{4}{*}{\shortstack{Vapor\\Pressure}} & \multirow{4}{*}{HQMLP} & \multirow{1}{*}{Noiseless}
        & HQFi/4Q & 0.9937 $\pm$ 0.0038 & 0.1600 $\pm$ 0.0432 & 0.8838 $\pm$ 0.0262 & 0.6896 $\pm$ 0.0345 \\
        \cmidrule(lr){3-8}
        & & \multirow{3}{*}{IBM-Fez}
        & HQSc/4Q & 0.9934 $\pm$ 0.0033 & 0.1786 $\pm$ 0.0541 & 0.8767 $\pm$ 0.0253 & 0.6869 $\pm$ 0.0185 \\
        & & & HQFi/4Q & 0.9952 $\pm$ 0.0017 & 0.1499 $\pm$ 0.0333 & 0.8836 $\pm$ 0.0240 & 0.6863 $\pm$ 0.0258 \\
        & & & HQFr/4Q & 0.9906 $\pm$ 0.0045 & 0.2424 $\pm$ 0.0654 & 0.8824 $\pm$ 0.0253 & 0.7214 $\pm$ 0.0395 \\
        \cmidrule(lr){1-8}
        \multirow{4}{*}{\shortstack{Boiling\\Point}} & \multirow{4}{*}{HQMLP} & \multirow{1}{*}{Noiseless}
        & HQFi/4Q & 0.9755 $\pm$ 0.0052 & 13.4047 $\pm$ 1.6366 & 0.8740 $\pm$ 0.0190 & 21.9557 $\pm$ 1.7384 \\
        \cmidrule(lr){3-8}
        & & \multirow{3}{*}{IBM-Fez}
        & HQSc/4Q & 0.9824 $\pm$ 0.0013 & 10.5738 $\pm$ 0.5862 & 0.8817 $\pm$ 0.0342 & 20.0476 $\pm$ 0.7922 \\
        & & & HQFi/4Q & 0.9776 $\pm$ 0.0064 & 13.3781 $\pm$ 3.7997 & 0.8687 $\pm$ 0.0302 & 22.1864 $\pm$ 2.2045 \\
        & & & HQFr/4Q & 0.9748 $\pm$ 0.0038 & 13.6211 $\pm$ 1.0202 & 0.8770 $\pm$ 0.0279 & 21.7611 $\pm$ 0.6974 \\
        \cmidrule(lr){1-8}
        \multirow{4}{*}{\shortstack{Melting\\Point}} & \multirow{4}{*}{HQGNN} & \multirow{1}{*}{Noiseless}
        & HQFi/4Q & 0.9147 $\pm$ 0.0319 & 23.8698 $\pm$ 3.4422 & 0.7715 $\pm$ 0.0220 & 35.8185 $\pm$ 1.0808 \\
        \cmidrule(lr){3-8}
        & & \multirow{3}{*}{IBM-Fez}
        & HQSc/4Q & 0.9162 $\pm$ 0.0119 & 24.9574 $\pm$ 1.3404 & 0.7638 $\pm$ 0.0307 & 35.1008 $\pm$ 1.1552 \\
        & & & HQFi/4Q & 0.9204 $\pm$ 0.0175 & 23.8559 $\pm$ 2.6269 & 0.7817 $\pm$ 0.0246 & 35.8391 $\pm$ 1.3810 \\
        & & & HQFr/4Q & 0.9304 $\pm$ 0.0135 & 22.3581 $\pm$ 2.1128 & 0.7552 $\pm$ 0.0423 & 35.4722 $\pm$ 1.1055 \\
        \bottomrule
    \end{tabular}
\end{sidewaystable}

\end{document}